\documentclass[twocolumn,pre,floats,aps,amsmath,amssymb]{revtex4-1}
\usepackage{graphicx}  
\usepackage{bm}   

\usepackage{dcolumn}
\PassOptionsToPackage{export}{adjustbox}
\usepackage{amsmath}
\usepackage{amsfonts,array}
\usepackage{braket,mleftright}
\usepackage{mathtools}
\usepackage{booktabs}
\usepackage{mhchem}
\usepackage[utf8]{inputenc}
\usepackage{makecell}
\usepackage[colorlinks=true,citecolor=blue]{hyperref}
\usepackage{float}
\usepackage{etoolbox}
\usepackage{verbatim}  
\usepackage{array}  
\newcolumntype{P}[1]{>{\centering\arraybackslash}p{#1}}  
\newcolumntype{M}[1]{>{\centering\arraybackslash}m{#1}}  
\usepackage{multirow}  
\usepackage{etoolbox,fontawesome}
\usepackage{hyperref} 
\AtBeginEnvironment{thebibliography}{\def\url#1{\href{#1}{\faExternalLink}}}
\usepackage[normalem]{ulem}   
\usepackage{lipsum}

\begin{document}
\newpage
\title{Pressure induced Lifshitz transition and anomalous crystal field splitting in AFeAs (A=Li/Na) Fe-based superconducting compounds: A first principles study}
\author{Soumyadeep Ghosh}
\email{soumyadeep@rrcat.gov.in}
\author{Haranath Ghosh} \thanks{corresponding author}
\email{hng@rrcat.gov.in}
\affiliation{Homi Bhabha National Institute, 2nd floor, Training School Complex, Anushakti Nagar, Mumbai 400094, India}
\affiliation{Human Resources Development Section, Raja Ramanna Centre for Advanced Technology, Indore 452013, India}
\date{\today}

\begin{abstract}
	\noindent {\it Abstract}: The effect of hydrostatic pressure on the electronic structures of iron pnictide superconducting compounds LiFeAs and NaFeAs through density functional theory based first principles studies are presented, using the predicted crystal structures at very high pressure by Zhang \emph{et~al.} \cite{111_hps_PCCP}. The orbital selective pressure induced modifications in the partial density of states and electronic band structures reveal mixed multi-band multi-orbital nature of these compounds, with energetically degenerate $ d_{xz} $/$ d_{yz} $ orbitals at ambient pressure. Due to larger hydrostatic pressures some of the electron/hole bands crosses the Fermi level, leading to significant topological modifications in Fermi surfaces known as Lifshitz transition. Interrelation between Lifshitz transition and superconductivity-an important subject matter of current research are described. Based on such interconnection the current study predicts the superconducting-$T_c$ in 111 compounds can not be raised at such high pressures. Wannier functions based electronic structure investigation reveal a relatively larger hybridization between the Fe-$ 3d $ and As-$ 4p $ orbitals at higher pressures, as an effect of which the orbital degeneracy of $ d_{xz} $ and $ d_{yz} $ orbitals are lifted. Different hybridization contributions in the crystal-field splitting are separated using the Wannier functions based formalism, by incorporating different bands with a particular orbital character in the Wannier function construction. Pressure dependence of the intra/inter orbital hopping amplitudes between Fe-$ d $ orbitals has been discussed using low energy tight binding model.

	\vspace{1pc}
	
	\noindent {\it Keywords}: Iron pnictides superconducting materials, Electronic structure calculations, Orbital characters, Lifshitz transition, Wannier functions, Crystal field splitting, Hopping parameters. 
\end{abstract}

\maketitle

\section{Introduction}
Microscopic origin of superconductivity in iron-based compounds (FeSCs) is still a mystery even after fourteen years of its discovery \cite{Kamihara26K}. One of the very important facts about FeSCs is the occurrence of Lifshitz like topological transition (LT). LT occurs at absolute zero temperature in which Fermi surface topology of metallic compounds change abruptly \cite{abyay_1144,abyay_12442,abyay_112}. No spontaneous symmetry breaking is involved in LT. LT is also observed in different condensed matter systems like \ce{Bi_2Sr_2CaCu_2O_{8+\delta}} cuprate superconductor \cite{LT_ab_th70}, 3D Dirac semi-metal \ce{Na_3Bi} \cite{LT_ab_th71}, bilayer graphene \cite{LT_ab_th72}, quantum hall liquids \cite{LT_ab_th73} \textit{etc}. FeSCs are multiple-band superconductors with specific orbital characters. Although typically FeSCs have multiple electron as well as hole like Fermi surfaces, but there are a number of materials which probably do not exhibit the hole Fermi surfaces. For example, \ce{Ba_{1-x}K_xFe_2As_2} near $x \sim 0.5$ \cite{LT_chem_doping}, \ce{LiFe_{1-x}Co_xAs} for $x \le 0.1$ \cite{LT_ab_th75}, K-dosed FeSe thin films \cite{LT_ab_th77,LT_ab_th78}, FeSe single layer grown on \ce{SrTiO_3} (FeSe/STO) substrate \cite{LT_ab_th76}, the intercalated compound (LiFe)OHFeSe \cite{LT_ab_th79} \textit{etc}. Due to multi band nature of electronic structure at the vicinity of the Fermi Level, it would be possible to tune the movement of any one or more number bands downward or upward depending on which high symmetry points in momentum space it crosses the Fermi Level through various external perturbations such as doping \cite{LT_chem_doping}, pressure \cite{LT_gupta} or even magnetic field \cite{LT_magnetic} causing Lifshitz transitions. In literature there exists a reasonably well established inter-connection between the LT and the highest superconducting critical transition temperature ($T_c$). For example, in \ce{Ba_{1-x}K_xFe_2As_2}, the concurrence of highest $T^{max}_c \sim 38$ K at $x\sim 0.50$ and an electronic topological transition occurs at the same doping concentration \cite{LT_ab_th89,LT_chem_doping}; similarly in \ce{BaFe_{2-x}Co_xAs_2}, a common doping value of $x\sim 0.10$ for the occurrence of $T^{max}_c \sim 26$ K and LT \cite{LT_ab_th92}; in \ce{BaFe2(As_{1-x}P_x)_2}, the occurrence of $T^{max}_c \sim 27$ K at $x\sim 0.35$ and LT at $x\sim 0.37$ \cite{LT_ab_th93}; in \ce{Ca_{0.82}La_{0.18}Fe_{1-x}Ni_xAs_2}, $T^{max}_c \sim 35$ K at $x\sim 0.35$ and LT occurs at same doping concentration \cite{LT_ab_th94}-are clear indications of existing interconnection between LT and $T_c$. So, FeSCs have become proto-type systems where probability of Lifshitz transition and its connection with superconductivity need to be studied. High pressure is one of the most versatile parameter which can tune $T_c$ as well as electronic properties of the system \cite{sen,soumya3}, though the study of pressure induced LT is rare. In this work we bring out the situation of pressure induced LT and its implication on superconductivity for two FeSC compounds LiFeAs and NaFeAs.

In general for high critical temperature superconductors, it is an well established fact that the pressure is a crucial controlling parameter of superconducting-$T_c$ and FeSCs are no exception of that \cite{quader}. For example, in case of superconducting compound $\beta$-\ce{Fe_{1.01}Se}, transition temperature is increased from 8.5 K to 36 K under the application of 8.9 GPa pressure \cite{11natmet2009}. One more example is \ce{BaFe_2As_2}, a parent FeSC which posses a spin-density-wave (SDW) ground state with no superconducting properties but hydrostatic/chemical pressure (by means of doping in any of the three atomic sites) can induce superconductivity with a $T_c \sim$  30 K \cite{mani_epl}. LiFeAs was the first discovered superconducting compound, belonging to 111 family with $T_c \sim$ 18 K \cite{LiFeAs_discovery}. The parent LiFeAs structure crystallizes in tetragonal symmetry with space group-$P4/nmm$, undergoes an structural phase transition under the application of hydrostatic pressure, as predicted by the theoretical study \cite{111_hps_PCCP}. It follows a structural transition sequence $ P4/nmm \rightarrow P\bar{3}m1 \rightarrow I4mm \rightarrow P6_3/mmc $ as predicted by the CALYPSO structure prediction method \cite{111_hps_PCCP} under the application of 40 GPa, 60 GPa and 240 GPa pressure respectively. NaFeAs, another parent member of 111 family, shows superconductivity around $T_c \sim$ 9 K, crystallize in primitive tetragonal structure with space group $P4/nmm$ \cite{NaFeAs_discovery}. The critical transition temperature of \ce{Na_{1-x}FeAs} poly-crystalline sample can be enhanced up to 31 K as the pressure increases from ambient pressure to 3 GPa \cite{zhang_epl}. In case of hole doped \ce{NaFe_{1-x}Co_xAs} (with x = 0.075) an enhancement of $T_c$ by 13 K is achieved under a pressure of 2.3 GPa \cite{wang}. According to the theoretical study \cite{111_hps_PCCP}, under external pressure (hydrostatic) NaFeAs show following structural transition in sequence $P4/nmm \rightarrow  Cmcm \rightarrow P\bar{3}m1 $. However, in the current literature the high pressure electronic structure and its implications on superconductivity are still missing. In this work, we provide a detailed systematic evolution of the electronic structures including crystal filed splitting with external pressure up to 240 GPa.

Crystal field splitting (CFS) is defined as the energy difference between the highest and the lowest orbital energy levels in the electron cloud of a transition metal element. Initially, the electrostatic potential of the surrounded ions was considered as the major source of CFS energy. But sooner, it became clear that hybridization between different orbitals of transition metal ions with the surrounded ligands must be considered for an accurate description of CFS \cite{WF_JPCM_r9}. Other factors affecting the CFS energy are, (i) symmetry and coordination environment of the transition metal ion, (ii) the valance state of the cation, (iii) pressure and temperature. Under perfect tetrahedral crystal field at ambient pressure, the five fold degenerate $ d $-orbitals split into the triply degenerate $ t_{2g} $ levels and doubly degenerate $ e_g $ levels. All the different family members of the FeSCs, share the similar Fe-As layer with tetrahedral symmetry at ambient pressure. Due to the effect of hybridization between the Fe $ d $-orbitals and the pnictogen/chalcogen-$ p $ orbitals both the $ t_{2g} $ and $ e_g $ orbitals further split, leaving only degenerate $ d_{xz} $ and $ d_{yz} $ orbitals \cite{de_medici}. Hydrostatic pressure can further modify this degree of hybridization, hence the corresponding crystal field splitting. Therefore, we have employed the maximally localized Wannier function (MLWF) formalism to analyze the effect of $ d-p $ hybridization due to applied pressure on crystal field splitting \cite{WF_JPCM,1144_cfs}.

From the above discussions, it is clear that both the electronic structure and CFS can be affected significantly  by the application of hydrostatic pressure. In this article, we present a systematic study of both the properties using Density functional theory (DFT) based first principles method. In Sec. \ref{computation} we will discuss about the detail computational methodologies. Sec. \ref{results} is subdivided into five parts, effect of pressure on: (\ref{sec_opbs}) orbital projected electronic band structure (BS) and Fermi surfaces, (\ref{sec_pdos}) orbital projected partial density of state (PDOS), (\ref{sec_cfs}) crystal field splitting in Wannier function formalism, (\ref{tb_Li}/\ref{tb_Na}) low energy tight binding models. Finally in Sec. \ref{conclusion}, we summarize the main conclusions of this work.

\section{Computational details} \label{computation}
The DFT-based electronic structure calculations are performed using the plane wave pseudo-potential based Quantum ESPRESSO code \cite{espresso}. The electronic exchange correlation is considered within the generalized gradient approximation (GGA) with Perdew-Burke-Enzerhof (PBE) functional \cite{Perdew}. Here we have used experimental lattice parameters of LiFeAs and NaFeAs at atmospheric pressure \cite{LiFeAs_discovery,NaFeAs_discovery} for the non spin polarized single point energy calculations. Zhang \textit{et al.} theoretically determines the crystal structures of 111-FeSC materials at high pressures, based on particle-swarm optimization algorithm \cite{111_hps_PCCP}. These structural details are used as inputs in our calculations for electronic structure at high pressures. The plane wave cut-off energy for self consistent field (SCF) calculations are taken appropriately, after performing a rigorous convergence test. Monkhorst-Pack scheme has been used to sample the Brillouin zone (BZ) in k-space \cite{Monkhorst,Pack}. We have employed the WANNIER90 package \cite{wannier90} to simulate low energy tight binding model and crystal field splitting. We use VETSA software package \cite{vesta} for visualization of Wannier functions and XCRYSDEN package \cite{xcrysden} for the visualization of Fermi surfaces. Throughout the article the Fermi energy level is set to zero eV.

\section{Results and Discussions} \label{results}
\begin{figure*}
	\includegraphics[width=15cm,height=14cm]{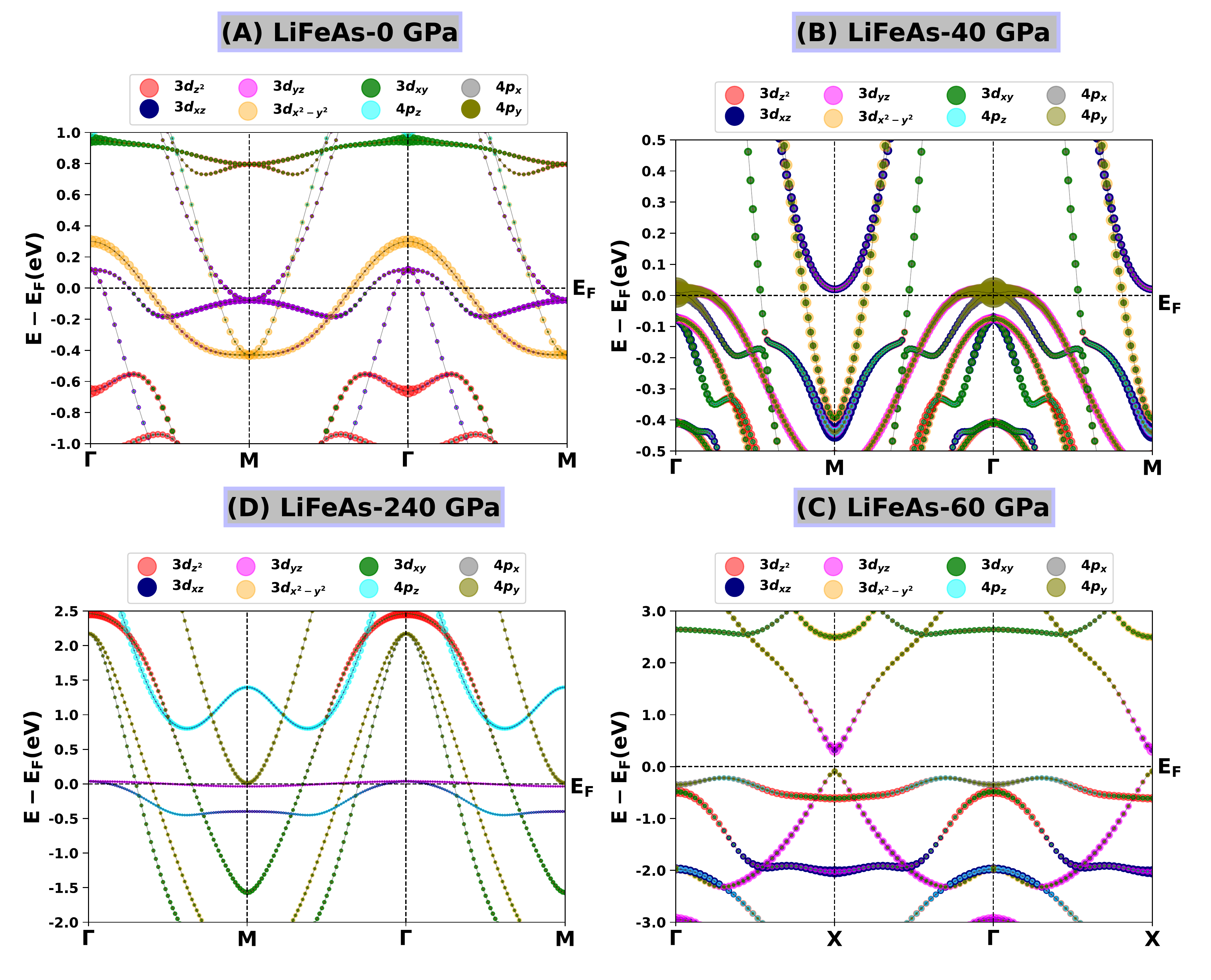}
	\caption{(Color online) Orbital projected electronic band structure of LiFeAs at various pressures. Atmospheric pressure: (A) 0 GPa,  high pressures: (B) 40 GPa, (C) 60 GPa, (D) 240 GPa. Fermi energy level is set to zero. Almost localized like $d_{yz}$ states are worth noticing at 240 GPa (cf. Fig. \ref{fig:LiFeAs-240GPa_OPBS_fl}).}
	\label{fig:LiFeAs_OPBS}
\end{figure*}

\subsection{Effect of pressure on Electronic band structure, Fermi surfaces and occurrence of Lifshitz like electronic topological transitions} \label{sec_opbs}

\begin{figure}
	\includegraphics[width=6cm, height=3cm]{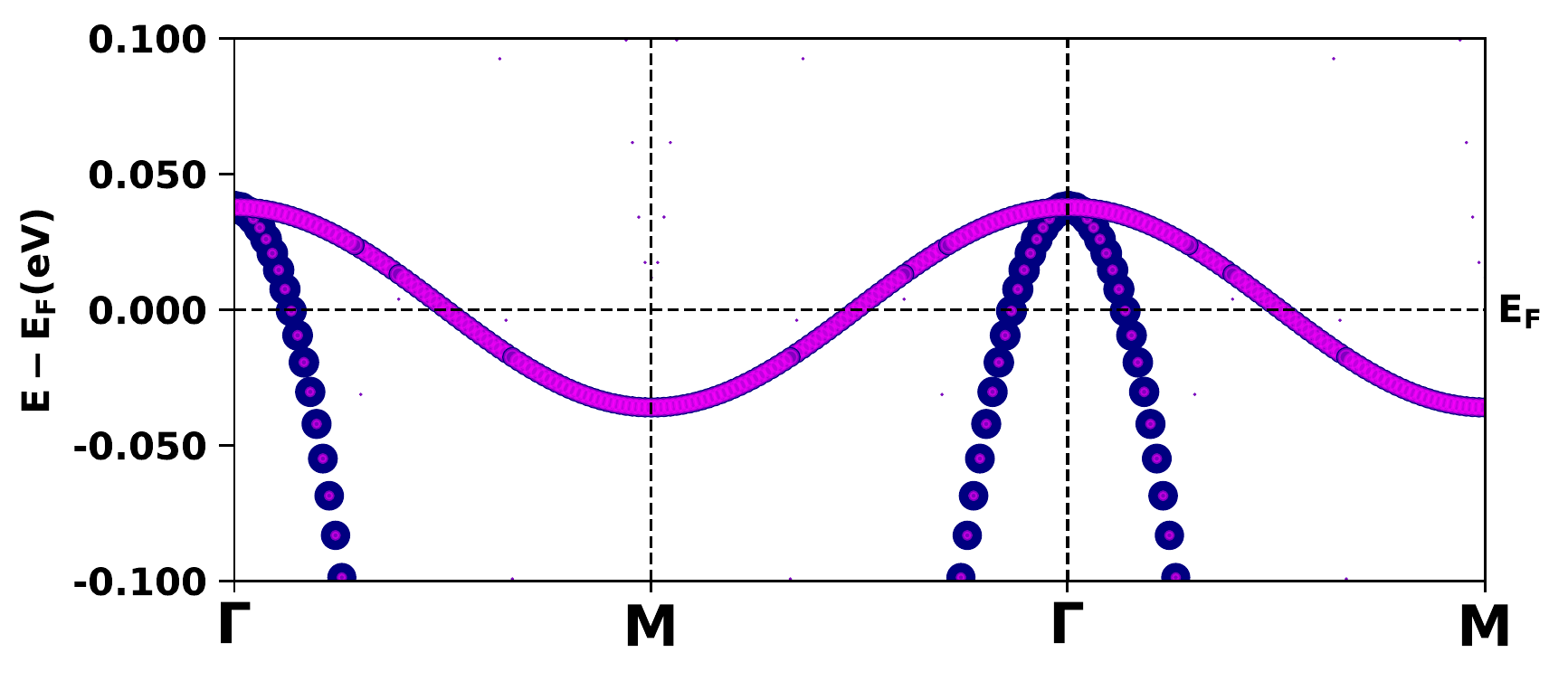}
	\caption{(Color online) Low energy part of the LiFeAs orbital projected band structure at 240 GPa. Only $ d_{xz} $ (blue) and $ d_{yz} $ (pink) orbitals are shown in the figure. Fermi energy level is set to zero.}
\label{fig:LiFeAs-240GPa_OPBS_fl}
\end{figure}

LiFeAs crystallizes in tetragonal crystal structure with space group $P4/nmm$ \cite{LiFeAs_discovery}. It shows structural transitions under the application of the hydrostatic pressure \cite{111_hps_PCCP}. It possesses trigonal, tetragonal and hexagonal crystal structures at 40 GPa, 60 GPa and 240 GPa pressure respectively. The orbital projected electronic band structures (OPBS) of LiFeAs at four different pressures are presented in Fig. \ref{fig:LiFeAs_OPBS}, around the high symmetry points. The different orbital contributions are denoted by assigning a color to each orbital involved in the band formation. The orbital weight to the bands is proportional to the size of the circles. Red, blue, pink, yellow, green, cyan, gray, and olive colors are assigned to signify the $ 3d_{z^2} $, $ 3d_{xz} $, $ 3d_{yz} $, $ 3d_{x^2-y^2} $, $ 3d_{xy} $, $ 4p_{z} $, $ 4p_{x} $, and $ 4p_{y} $ orbitals, respectively. We have followed this same color scheme throughout this work. We have projected different atomic orbitals onto Kohn-Sham states to calculate the orbital character of different bands. OPBS reveal mixed multi-orbital multi-band nature in all the four different crystallographic symmetries studied here. At atmospheric pressure LiFeAs posses two electron like bands around M point and three hole like bands around \ce{\Gamma} point \cite{LiFeAs_bs_prl} near Fermi level (see Fig-\ref{fig:LiFeAs_OPBS}A). Out of the three hole like bands around \ce{\Gamma} point, one band crosses the Fermi level (FL), having $ d_{x^2-y^2} $ orbital character while other two bands lie just above the FL are almost degenerate, having mostly mixed $ d_{yz} $, $ d_{xy} $ orbital character. There are two electron like bands that crosses the FL, of which one lies close to the FL and the other one is away from the FL. The band close to the FL is $ d_{yz} $ orbital derived and the band away from the FL is primarily of Fe $ d_{x^2-y^2} $, As $ 4p_z $ orbital character. When pressure increases to 40 GPa, there is a drastic change in the electronic band structure that modifies to two hole like bands around \ce{\Gamma} point and two electron like bands around M point (see Fig-\ref{fig:LiFeAs_OPBS}B). Change in the orbital characters of the bands are also noteworthy. Two hole like bands are almost degenerate near the FL, one having As $ 4p_y $, $ d_{yz} $ orbital character while the other is As $ 4p_y $, $ d_{xy} $ orbital derived. Whereas both the electron like bands lie away from the FL, one having $ d_{x^2-y^2} $, $ d_{xy} $ orbital characters while the other has mixed $ d_{xz} $, $ d_{yz} $, $ d_{xy} $, As $ 4p_z $ character. It is worthwhile pointing out that both the hole like bands near $\Gamma$ point and the $d_{xz}$ orbital derived electron like band at M point are on the verge of LT. A large number of hole like bands have already crossed the FL at $\Gamma$ and M points, such a behaviour is equivalent to large electron doping. At 60 GPa, LiFeAs experiences a metal-semiconductor transition where both the valance bands and conduction bands have multi-orbital nature (see Fig-\ref{fig:LiFeAs_OPBS}C). Valance bands near the FL at \ce{\Gamma} point have mixed $ d_{z^2} $, $ d_{xy} $ and As $ 4p_y $ character. Conduction bands as well as valance bands at X point have mixed multi-orbital character comprising of $ d_{yz} $, $ d_{xz} $ and As $ 4p_y $. In case of 240 GPa pressure, we observe five hole like bands at \ce{\Gamma} point and one electron like band at M point (see Fig-\ref{fig:LiFeAs_OPBS}D). Among the five hole like bands two of them lie very close to the FL and other three lie away from the FL. These two lower lying bands are almost degenerate and mostly have mixed $ d_{xz} $, $ d_{yz} $, As $ 4p_z $ orbital character. Among three upper lying bands two are almost degenerate and mostly of As $ 4p_y $ derived, while the third one is primarily $ d_{z^2} $ orbital derived. The only electron like that band lies near FL is mostly derived from Fe $ d_{yz} $ orbital (see Fig-\ref{fig:LiFeAs-240GPa_OPBS_fl}). Therefore, larger presence of As-$ p $ orbital character bands near the FL suggest that, at relatively higher pressure the mixing between Fe-$ 3d $ and As-$ 4p $ orbital increases. Another most important observation is, degeneracy of $ d_{xz} $ and $ d_{yz} $ orbitals at ambient pressure is lifted due to applied hydrostatic pressure. 

\begin{figure}
	\includegraphics[width=9.3cm,height=12cm]{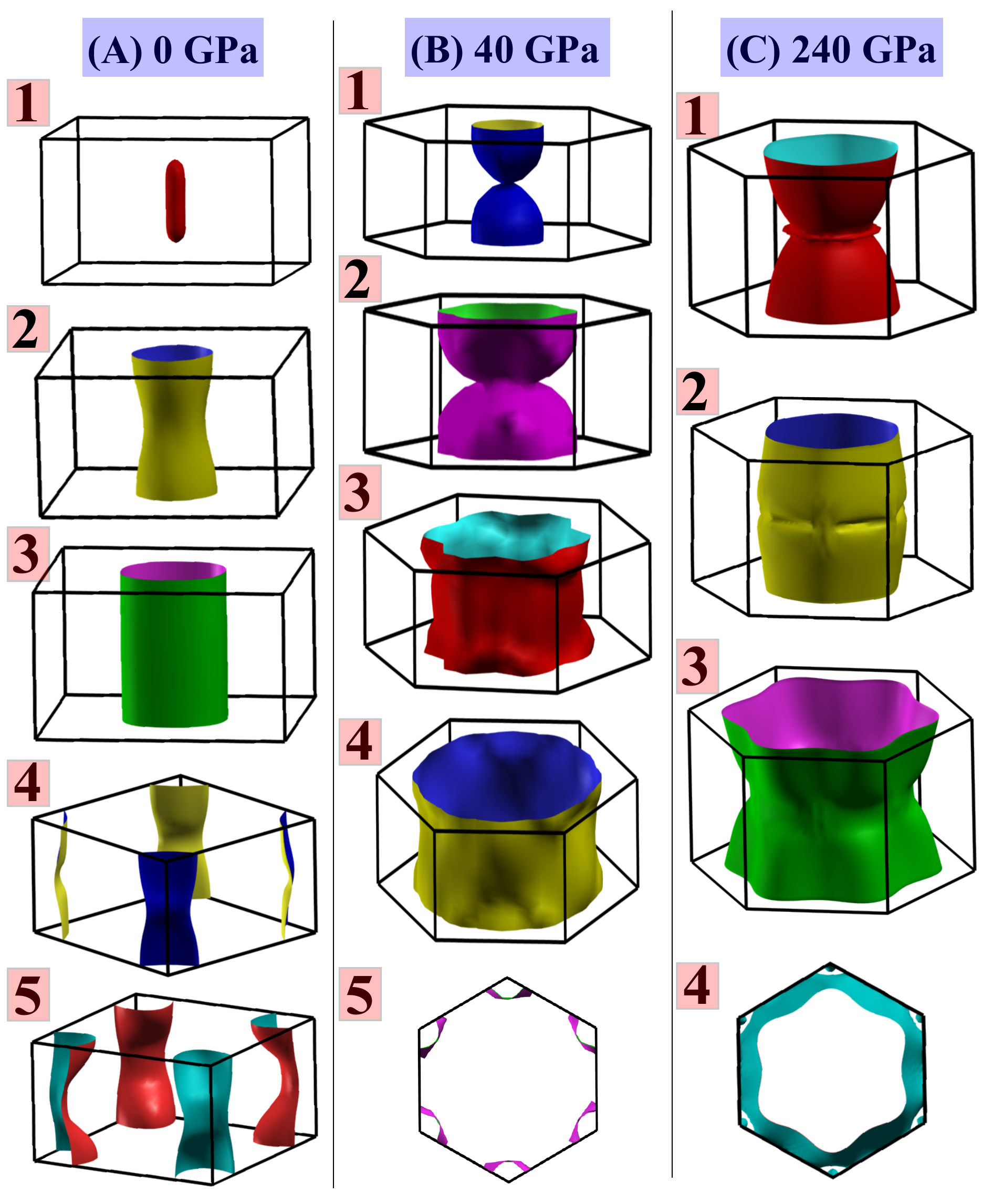}
	\caption{(Color online) Fermi surfaces of LiFeAs at various pressures. Atmospheric pressure: (A) 0 GPa,  high pressures: (B) 40 GPa, and (C) 240 GPa. Xcrysden package is used for visualization \cite{xcrysden}.}
	\label{fig:LiFeAs_FS}
\end{figure}

\begin{figure*}
	\includegraphics[width=19cm,height=12cm]{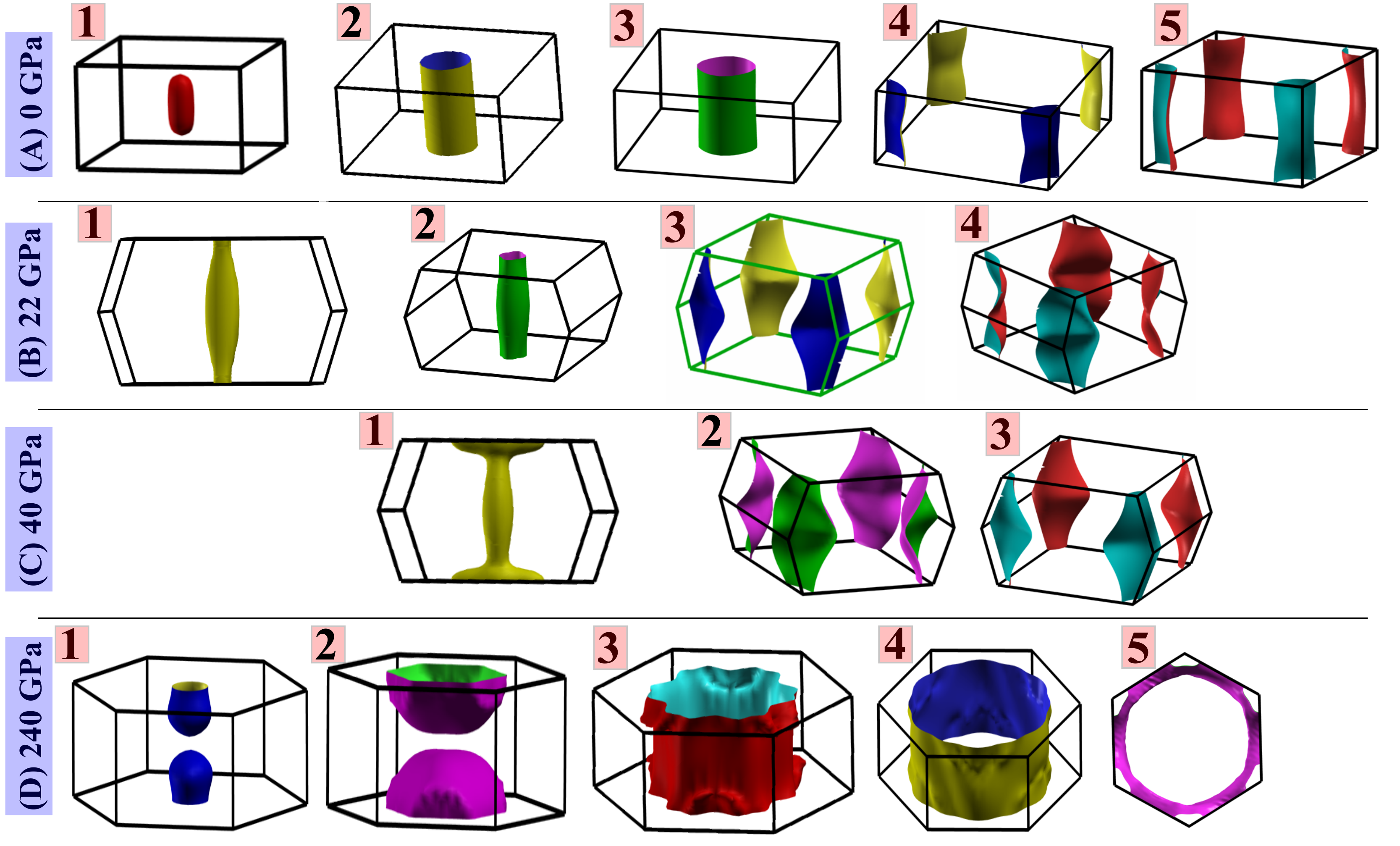}
	\caption{(Color online) Fermi surfaces of NaFeAs at various pressures. Atmospheric pressure: (A) 0 GPa,  high pressures: (B) 22 GPa, (C) 40 GPa, and (D) 240 GPa. Xcrysden package is used for visualization \cite{xcrysden}.}
	\label{fig:NaFeAs_FS}
\end{figure*}

\begin{figure*}
	\includegraphics[width=16cm,height=7cm]{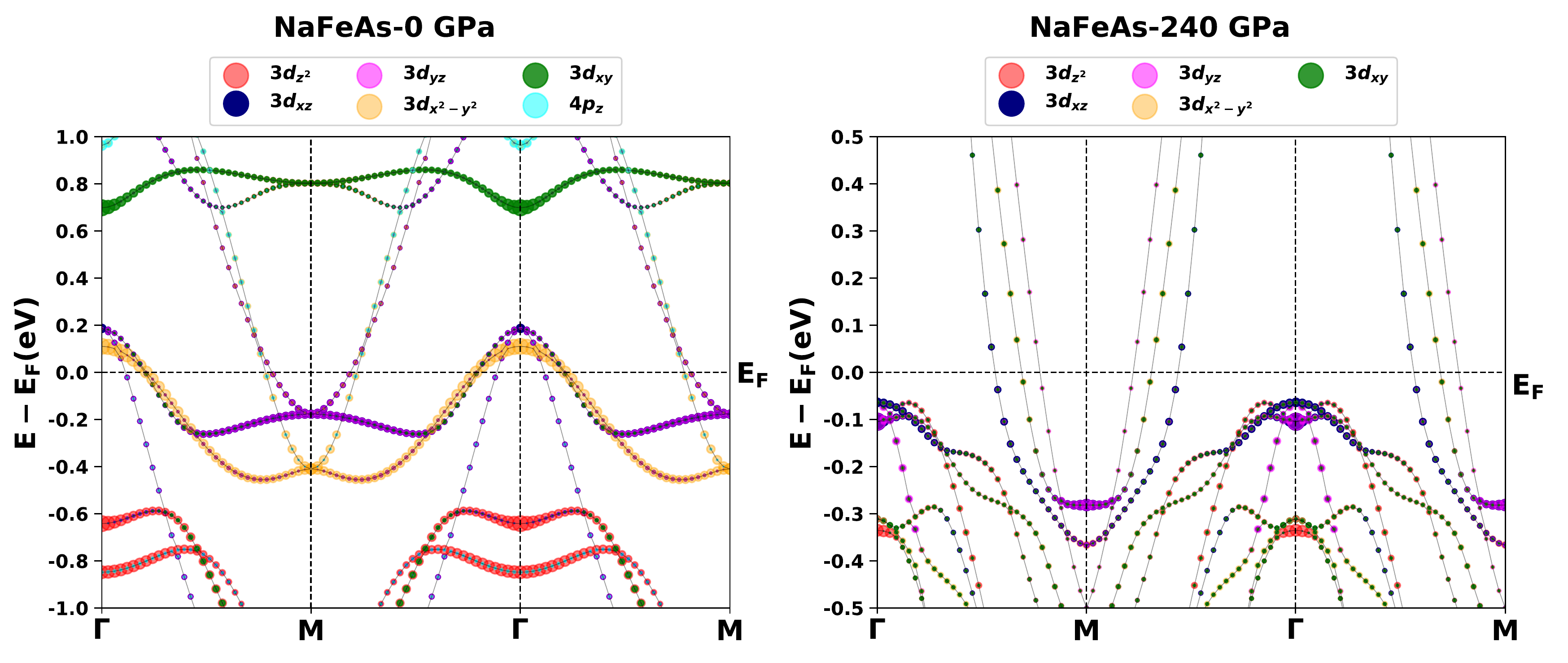}
	\caption{(Color online) Orbital projected band structure of NaFeAs at ambient pressure and 240 GPa pressure. Fermi energy level is set to zero. No hole like Fermi surface is expected at 240 GPa due to pressure induced Lifshitz transition. This is in contrast to LiFeAs (cf. Fig. \ref{fig:LiFeAs_OPBS}D).}
	\label{fig:Na-OPBS_0_240}
\end{figure*}

In case of iron based FeSC compounds the discussions regarding topology of Fermi surface (FS) is very much important. It may reveal various crucial information like possibility of Lifshitz like topological transitions, nesting condition etc. \cite{abyay_intermetallics}. Here, we will discuss about the evolution of FSs in LiFeAs compound at various pressures. At ambient pressure (Fig. \ref{fig:LiFeAs_FS}A) there are three hole like FSs at the Brillouin zone centre (around $\Gamma$-point) whereas there are two electron like FSs at the Brillouin zone corner (around $M$-point). At 40 GPa pressure (Fig. \ref{fig:LiFeAs_FS}B) there are four hole like FSs at the Brillouin zone centre and only one electron like FS at the Brillouin zone corner. Similar kind of hole/electron like FSs are found in first hexagonal Brillouin zone of hexagonal \ce{MgB_2} \cite{hexagonal_FS_MgB2}/other 2D-metallic compounds \cite{hexagonal_FS_2d} respectively. At 240 GPa pressure (Fig. \ref{fig:LiFeAs_FS}C) there are three hole like FS at the Brillouin zone centre and only one electron like FS at the Brillouin zone corner.  Interestingly, the electron like FS has a very different character, consisting of rings connected by bridges and topologically parallel to the outside faces of the Brillouin zone. Similar kind of FSs are found in hexagonal FeS polymorph \cite{hexagonal_FS_FeS}. So, hydrostatic pressure influences modification in topology of both the hole and electron like FSs. At ambient pressure at $\Gamma$-point the hole FSs look like concentric cylinders, but with increasing pressure the radius of cylinder increases as well as slightly distorted in their shape. So, the nesting between hole like FS and electron like FS may be possible. This indicates enhanced possibility of repulsive inter band pairing interaction between hole and electron like FSs which may be advantageous for $s^\pm$ symmetry of the superconducting gap \cite{abyay_12442}. As the inner electron pockets are on the verge of LT, it is also possible that the nesting between hole and electron pockets are relatively weak. In Fig. \ref{fig:LiFeAs_FS} hole FS area increases with pressure whereas electron FSs undergo missing is an example of pressure induced LT. Here, our discussion about the nesting is qualitative only, explicit calculation of spin susceptibility/nesting function is required to comment on nesting property rigorously and quantitatively \cite{nesting}.

\subsubsection*{Fermi surfaces of NaFeAs and occurrence of Lifshitz transition}

At ambient pressure NaFeAs also crystallizes in tetragonal crystal structure \cite{NaFeAs_discovery}. When hydrostatic pressure is applied, it shows several structural transitions \cite{111_hps_PCCP}. It possesses orthorhombic crystal structure upto pressures 40 GPa, then transformed to trigonal crystal structure. FS topology of NaFeAs at various pressures is shown in Fig. \ref{fig:NaFeAs_FS}. At ambient pressure (Fig. \ref{fig:NaFeAs_FS}A) there are three hole like FS at the Brillouin zone centre (around $\Gamma$-point) whereas there are two electron like FS at the Brillouin zone corner (around $M$-point). At 22 GPa pressure (Fig. \ref{fig:NaFeAs_FS}B) there are two hole like FSs at the Brillouin zone centre and two electron like FSs at the Brillouin zone corner. At 40 GPa pressure (Fig. \ref{fig:NaFeAs_FS}C) there are only one hole like FS at the Brillouin zone centre and two electron like FSs at the Brillouin zone corner. So, the shape and size of hole like FS are changing drastically with the applied pressure. Similar kind of hole like FSs are found in first hexagonal Brillouin zone of anti-ferromagnetic \ce{BaFe_2As_2} \cite{hexagonal_FS_PRL}. At 240 GPa pressure (Fig. \ref{fig:NaFeAs_FS}D) the innermost hole like FS segregates into two parts leading to Lifshitz transition. The same can be inferred from electronic band structure at Fig. \ref{fig:Na-OPBS_0_240}. The hole like bands at $\Gamma$-point gradually shifts below the FL with applied pressure. Topological modifications with appearance or destruction of new FS are the primary signatures of LT. Thus it is an example of hydrostatic pressure induced LT in NaFeAs. Again LT restricts superconducting $T_c$ \cite{LT_chem_doping,LT_gupta}, hence it could not be raised further with increasing pressure due to occurrence of LT.

\begin{figure*}
	\includegraphics[width=15cm,height=14.5cm]{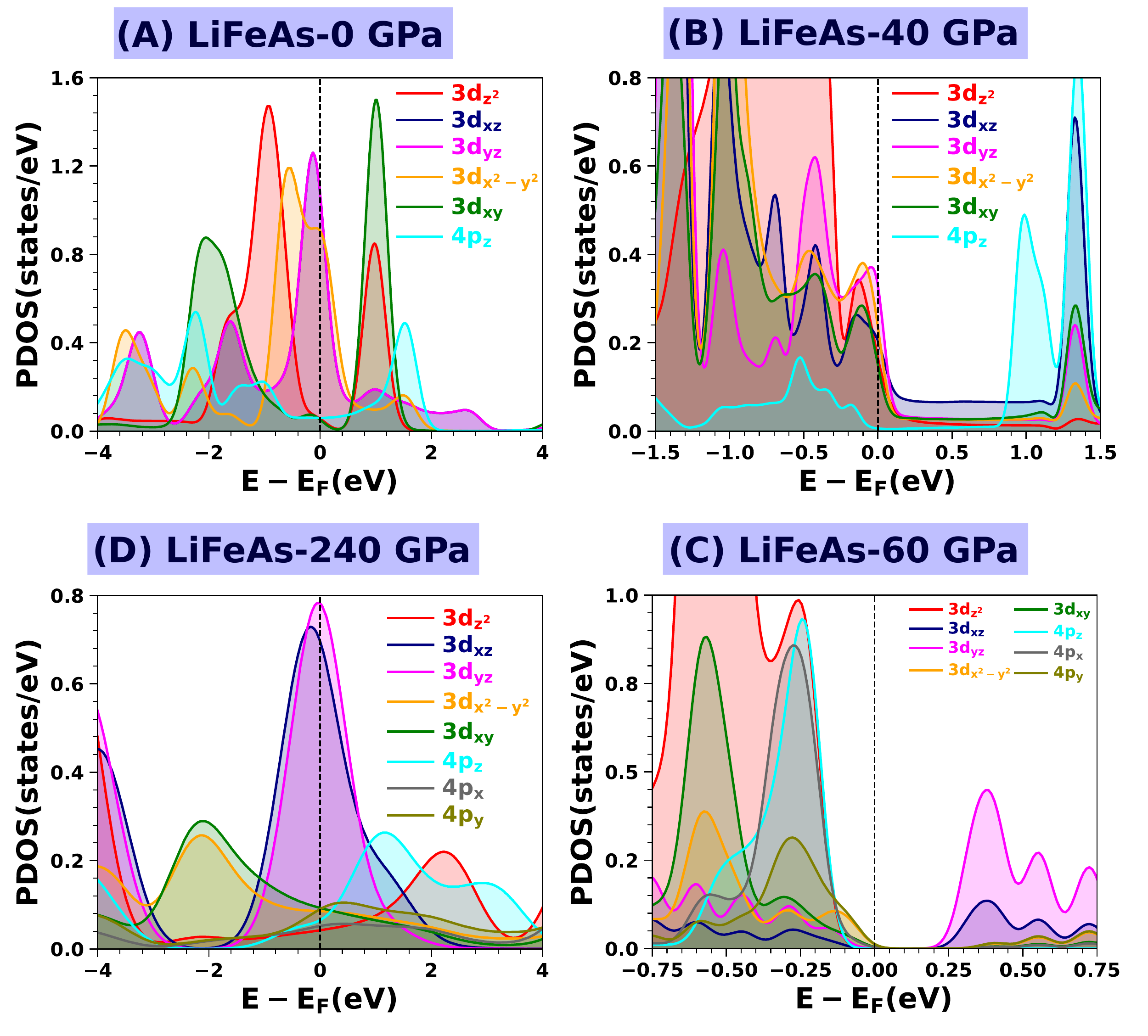}
	\caption{(Color online) Orbital selective pressure induced modifications in the PDOS of LiFeAs at various pressures. Atmospheric pressure: (A) 0 GPa,  high pressures: (B) 40 GPa, (C) 60 GPa, (D) 240 GPa. Fermi energy level is set to zero.}
	\label{fig:LiFeAs_PDOS}
\end{figure*}

\subsection{Pressure dependent electronic structure of LiFeAs and NaFeAs} \label{sec_pdos}
\subsubsection*{Orbital projected PDOS of LiFeAs}
\begin{figure*}
	\includegraphics[width=15cm,height=14cm]{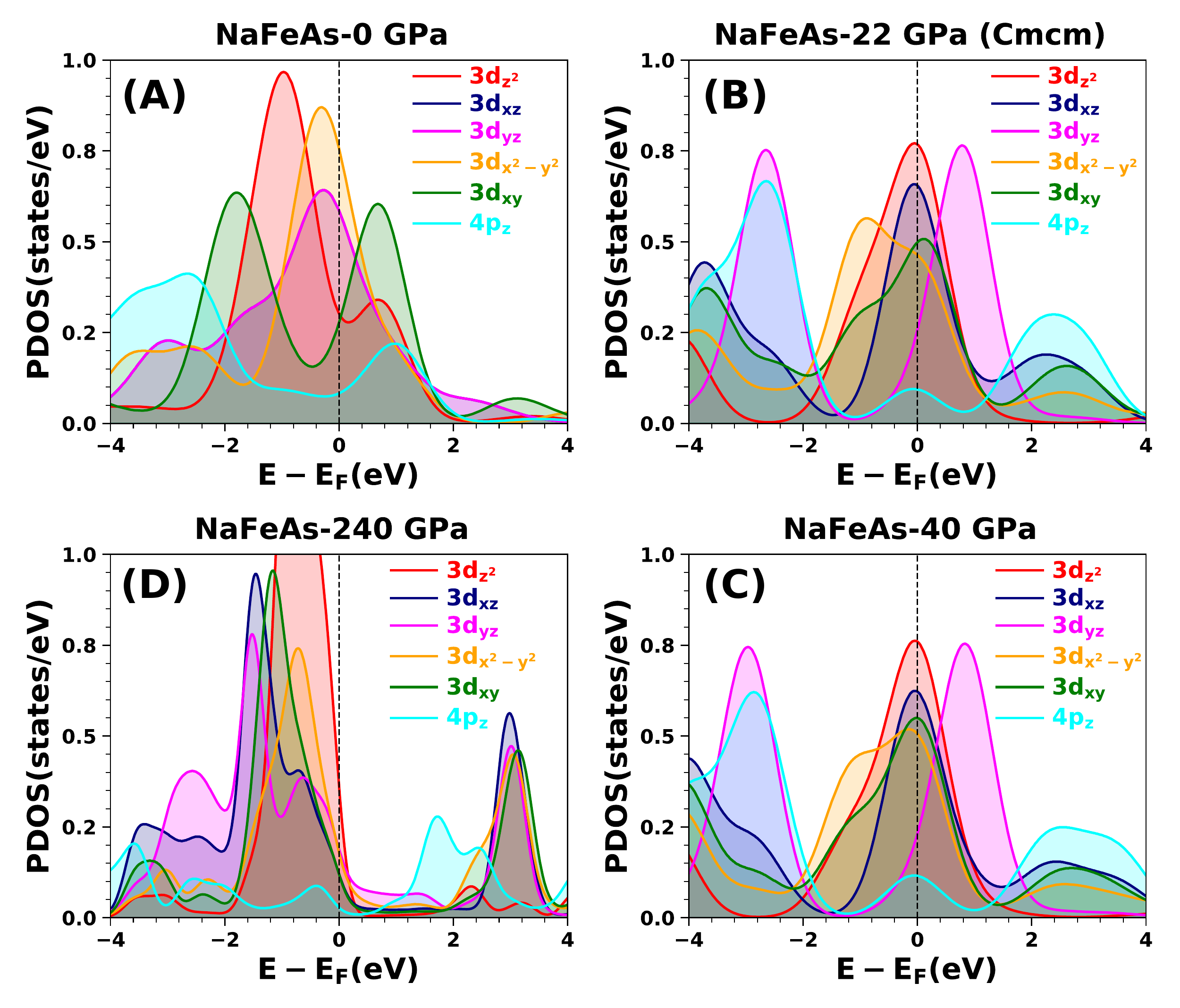}
	\caption{(Color online) Orbital selective pressure induced modifications in the PDOS of NaFeAs at various pressures. Atmospheric pressure: (A) 0 GPa,  high pressures: (B) 22 GPa, (C) 40 GPa, and (D) 240 GPa. Fermi energy level is set to zero.}
	\label{fig:NaFeAs_PDOS}
\end{figure*}

In Fig. \ref{fig:LiFeAs_PDOS} partial density of states of Fe-$ 3d $ and As-$ 4p $ orbitals for LiFeAs at different pressures are presented. The maximum contribution in the PDOS around the FL arises from the Fe-$ 3d $ orbitals except at 60 GPa pressure, where As-$ 4p $ orbitals also have significant contribution in the valance states. In case of LiFeAs, at atmospheric pressure the maximum contribution to the PDOS at FL arises from the $ d_{yz} $ and $ d_{x^2-y^2} $ orbitals, out of five Fe-$ 3d $ orbitals (see Fig. \ref{fig:LiFeAs_PDOS}A). At atmospheric pressure the $ d_{yz} $ and $ d_{xz} $ orbitals contribute equally in the PDOS, hence they are degenerate in nature. The contributions of $ d_{z^2} $, $ d_{xy} $ and As $ 4p_{z} $ orbitals to the PDOS at FL are the lowest. At 40 GPa pressure, maximum contributions to the PDOS at FL also arises from the $ d_{yz} $ and $ d_{x^2-y^2} $ orbitals (see Fig. \ref{fig:LiFeAs_PDOS}B). Here the contributions from $ d_{xz} $, $ d_{z^2} $, $ d_{xy} $ orbitals are slightly less compared to other $ d $-orbitals, while As $ 4p_{z} $ orbital contribution is almost zero. Most importantly, at 40 GPa pressure we notice different $ d_{xz} $ and $ d_{yz} $ orbital contributions to the PDOS at FL. At 60 GPa pressure, a band gap of around 0.16 eV near FL opens up which splits the bands into valance and conduction bands (see Fig. \ref{fig:LiFeAs_PDOS}C). At valance band, maximum contributions to the PDOS arises from the $ d_{z^2} $, As-$ 4p_z $ and As-$ 4p_x $ orbitals. However, at conduction band $ d_{yz} $ and $ d_{xz} $ orbitals contribute the most in the PDOS. Here we observe $ d_{yz} $ and $ d_{xz} $ orbital contributions are also dissimilar. At 240 GPa pressure, maximum contribution to the PDOS at FL arises from the $ d_{yz} $ and $ d_{xz} $ orbitals (see Fig. \ref{fig:LiFeAs_PDOS}D). In contrast, $ d_{xy} $, $ d_{x^2-y^2} $ and As $ 4p_y $ orbitals contribute less but almost equally to the PDOS near FL. Here we also found a less but non zero contribution from the As $ 4p_z $ and $ d_{z^2} $ orbitals. Therefore, discussions on the overall nature of PDOS near the FL signals to the fact that individual Fe-$ 3d $ and As-$ 4p $ orbitals are affected differently due to the applied pressure. The above mentioned behavior of the $ d_{xz} $ and $ d_{yz} $ orbital PDOS demonstrate that, at relatively higher pressure they become non-degenerate.

\subsubsection*{Orbital projected PDOS of NaFeAs}
Orbital projected PDOS of Fe-$ 3d $ and As-$ 4p $ orbitals at different pressures are presented in Fig. \ref{fig:NaFeAs_PDOS}. In all the cases the maximum contributions to the PDOS near the FL are coming from Fe $ d $-orbitals. At ambient pressure maximum contributions to the PDOS at FL arises from $ d_{x^2-y^2} $ and $ d_{yz} $ orbitals (see Fig. \ref{fig:NaFeAs_PDOS}A); whereas $ d_{z^2} $ and $ d_{xy} $ orbital contributions are equal, but larger as compared to As $ 4p_z $ orbitals. From Fig. \ref{fig:NaFeAs_PDOS}A it is clear that $ d_{xz} $ and $ d_{yz} $ orbitals are degenerate at ambient pressure. This orbital degeneracy is lifted when hydrostatic pressure is applied (see Fig. \ref{fig:NaFeAs_PDOS}B). Here the maximum contributions to the PDOS at FL arises from $ d_{z^2} $ and $ d_{xz} $ orbitals. The contributions from $ d_{x^2-y^2} $ and $ d_{xy} $ orbitals are almost equal, but less as compared to $ d_{xz} $ orbitals. Here we also find less but non-zero contributions of $ d_{yz} $ and As $ 4p_z $ orbitals. At 40 GPa, $ d_{z^2} $ and $ d_{xz} $ orbital PDOS at FL slightly decreases, while $ d_{x^2-y^2} $ and $ d_{xy} $ orbital PDOS increases (see Fig. \ref{fig:NaFeAs_PDOS}C). Here we also observe slight decrease in PDOS of $ d_{yz} $ orbitals, whereas an increase in As-$ 4p_z $ orbital PDOS. At 240 GPa, largest contribution in PDOS near FL arises from Fe $ d_{z^2} $ orbitals. The contribution from other orbitals in decreasing order can be viewed as $ d_{x^2-y^2} $, $ d_{yz} $, $ d_{xz} $, $ d_{xy} $ (see Fig. \ref{fig:NaFeAs_PDOS}D). Here As $ 4p_{z} $ orbital contribution is negligible. Therefore, the outcome of above discussions is that, overall nature of Fe-$ 3d $ and As-$ 4p $ orbitals PDOS near FL are affected differently due to the applied pressure. PDOS at FL reduces at 240 GPa pressure, this in turn reduce the possibility of electron pairing at higher pressure, which may result in decreasing superconducting $T_c$. The observed nature of $ d_{xz} $ and $ d_{yz} $ orbitals PDOS in both the compounds (LiFeAs and NaFeAs) indicates that, at relatively higher pressure they may become non-degenerate. The same can be inferred from the electronic band structures. To physically interpret this degeneracy lifting mechanism, we have performed crystal field splitting calculation using Wannier function based formalism (Sec-\ref{sec_cfs}).

\subsection{Anomalous crystal field splitting} \label{sec_cfs}
In the presented electronic structure calculation (section-A/B) we have considered the Bloch states as the basis of the single particle electronic Hamiltonian. Bloch states in the periodic solids are characterized by a band index $ m $ and a wave vector $ \bm{k} $ in the first Brillouin zone (FBZ). The many body Hamiltonian is diagonal in this basis \cite{WF_JPCM}:

\begin{equation}
H=\sum_{m,\bm{k}} \epsilon_{m \bm{k}} \hat{b}^{\dagger}_{m \bm{k}} \hat{b}_{m \bm{k}}
\end{equation}

Here, $ \hat{b}^{\dagger}_{m \bm{k}} $ is the particle creation operator in the Bloch state $ \ket{\phi_{m \bm{k}}} $, $ \hat{b}_{m \bm{k}} $ is the particle destruction operator and $ \epsilon_{m \bm{k}} $ is the single particle energy of the corresponding state. The Bloch states are transformed into a set of atomic like Wannier functions (WFs) using a unitary transformation defined as \cite{WF_review}:

\begin{equation}
\ket{W_{\alpha \bm{R}}} = \frac{V}{(2\pi)^3} \int_{FBZ} d^3\bm{k} e^{-i\bm{k}\cdot\bm{R}} \sum_{m} U^{(\bm{k})}_{m\alpha} \ket{\phi_{m \bm{k}}}
\label{eq:WF}
\end{equation}

The Wannier functions are characterized by a unit cell index $ \bm{R} $ along with an additional index $ \alpha $ which discriminate different Wannier orbitals in the same unit cell. The $ \bm{k} $ dependent unitary matrix $ U^{(\bm{k})} $ mixes various Bloch functions at the same $ \bm{k} $ point to generate the WFs. But the Wannier functions generated using eq. (\ref{eq:WF}) are non-unique, because different sets of $ U^{(\bm{k})} $ leads to the different sets of Wannier orbitals. To overcome these difficulties, the Maximally localized Wannier functions (MLWF) technique is the most suitable way to define a unique set of Wannier functions \cite{WF_MLWF}. The corresponding many body Hamiltonian becomes non-diagonal in this basis:

\begin{equation}
H=\sum_{\alpha,\beta,\bm{R},\bm{R'}} h_{\alpha \bm{R},\beta \bm{R'}}  \hat{w}^{\dagger}_{\alpha \bm{R}} \hat{w}_{\beta \bm{R'}}
\end{equation} 

and the corresponding matrix element can be written as:

\begin{align}
h_{\alpha \bm{R},\beta \bm{R'}} &= \frac{V}{(2\pi)^3} \int_{FBZ} d^3\bm{k} e^{-i\bm{k}\cdot\bm{(R-R')}} \nonumber\\
&~~~~~~~~~~~~~~~~~~~~~~   \sum_{m} (U^{(\bm{k})}_{m\alpha})^*\epsilon_{m \bm{k}}U^{(\bm{k})}_{m\beta}
\label{eq:hopping}
\end{align}

In case of iron based superconducting compounds, at atmospheric pressure each MLWF is located on a specific atomic sites (Wannier centers) and has a clear orbital character. This allows us to interpret MLWFs as effective tight binding (TB) orbitals \cite{koepernik_PRB,abyay_12442, nakamura_long,abyay_12442_2}. If either $ \bm{R \ne R'} $ or the indices $ \alpha $ and $ \beta $ corresponds to Wannier orbitals $ n $ and $ n' $ at different atomic sites $ i $ and $ j $, the matrix element $ h_{\alpha \bm{R},\beta \bm{R}} $ in eq. (\ref{eq:hopping}) can be interpreted as the hopping amplitude between $ ni\bm{R} $ and $ n'j\bm{R'} $, usually denoted as $ t_{ni\bm{R},n'j\bm{R'}} $. If $ \alpha $ and $ \beta $ corresponds to same site index $ i $ and same orbital index $ n $ in the same unit cell $ \bm{R}=\bm{R'} $, then the corresponding matrix element $ h_{\alpha \bm{R},\beta \bm{R}} $ is called onsite energies ($ \epsilon_{ni} $). In this TB basis the onsite energy differences between different orbitals with predominant orbital character (\textit{e.g.,} Fe-$ d $ or As-$ p $) can be comprehended as crystal field splitting. 

Different sets of bands, with a predominant orbital characters are used to construct different sets of Wannier functions (depending on which $ m $ bands are included in the summation of eq. (\ref{eq:WF})). Using this feature we can construct three different sets of Wannier functions (TB models) corresponding to : (i) `d-model'-only effective Fe-$ d $ orbitals, (ii) `dp-model' containing Fe-$ d $ orbitals and As-$ p $ orbitals, and (iii) `dps-model' containing Fe-$ d $ orbitals, As-$ p $ orbitals, Li/Na-$ s $ orbitals. The onsite energy difference, obtained using different orbital projections, can be interpreted as decomposition of the total crystal field splitting \cite{WF_JPCM}.

\subsubsection*{Crystal field splitting of LiFeAs under pressure} 
\begin{figure*}
		\includegraphics[width=0.9\textwidth, height=12cm]{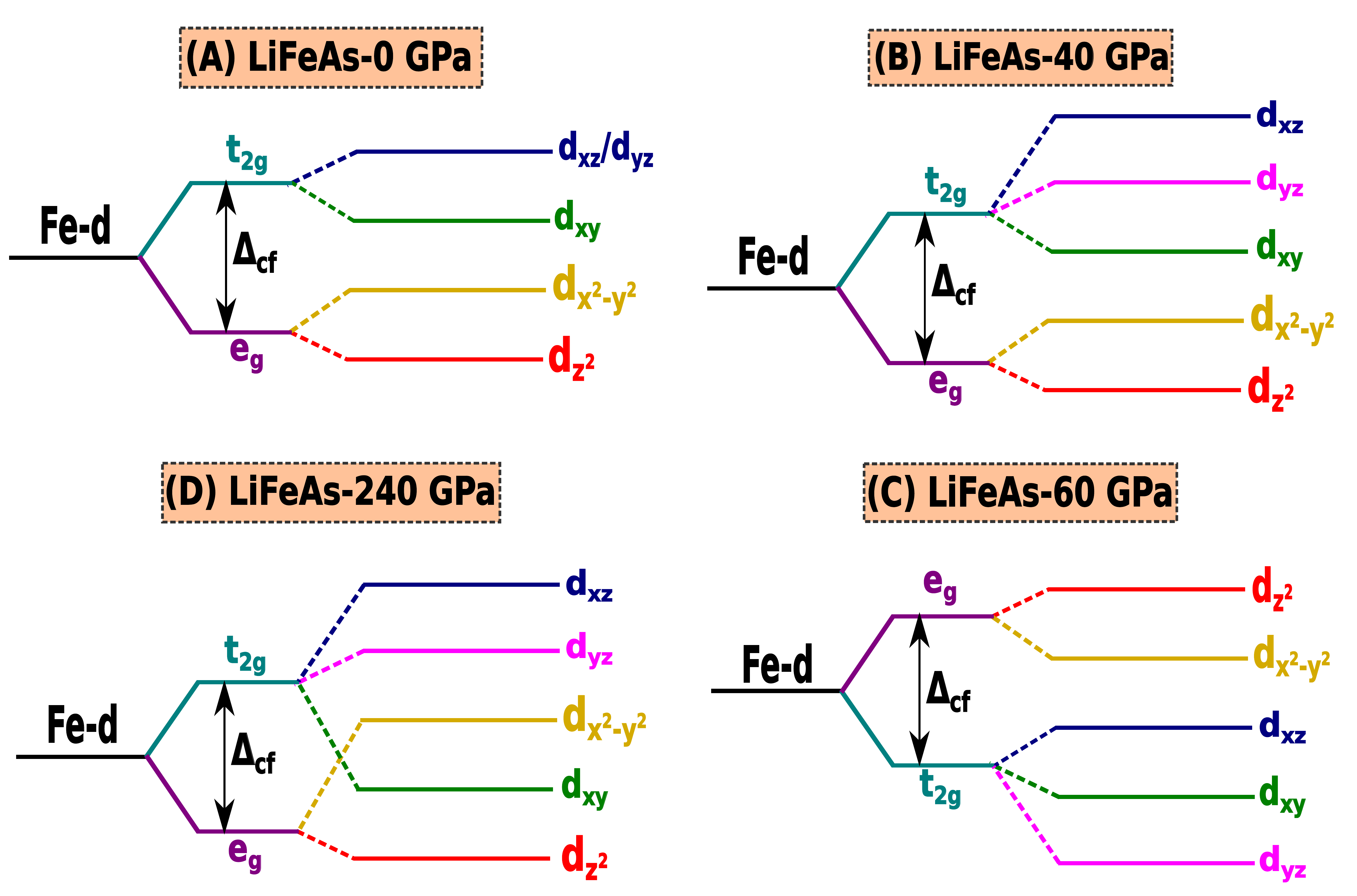}
		\caption{(Color online) Crystal field splittings of LiFeAs at various pressures (cartoon figure, not drawn to scale). Atmospheric pressure: (A) 0 GPa,  high pressures: (B) 40 GPa, (C) 60 GPa, (D) 240 GPa.}
		\label{fig:CFS_LiFeAs}
\end{figure*}

However, from Fig. \ref{fig:LiFeAs_OPBS} it can be seen that bands with predominant Fe-$ d $ and As-$ p $ orbital characters have major contributions in the electronic band structure near the FL. We construct MLWF for each group of bands separately and the resulting MLWFs can be viewed as atomic orbitals arising from hybridization between Fe/As/Li atoms. The onsite energies for the Fe-centered Wannier functions corresponding to the set of Fe-$ d $ orbitals are denoted as $ \epsilon^{(d)} $. The crystal field splitting using this so called `d-model' for LiFeAs at different pressures are presented in Fig. \ref{fig:CFS_LiFeAs}. At atmospheric pressure, in tetrahedral coordination, Fe-$ d $ orbitals splits into the $ t_{2g} $ $(d_{xz}, d_{yz}, d_{xy})$ and $ e_{g} $ $ (d_{x^2-y^2}, d_{z^2}) $ orbitals \cite{de_medici}. Here we found $ t_{2g} $ orbitals are lower in energy than that of the corresponding $ e_g $ orbitals (see Fig. \ref{fig:CFS_LiFeAs}A). The $ d_{xz} $ and $ d_{yz} $ orbitals are also degenerate. At 40 GPa pressure, in octahedral coordination, $ t_{2g} $ orbitals splits into non-degenerate $ d_{xz}, d_{yz}, d_{xy} $ orbitals and $ e_{g} $ orbitals splits into $ d_{x^2-y^2}, d_{z^2} $ orbitals (see Fig. \ref{fig:CFS_LiFeAs}B). Most importantly $ d_{xz} $ and $ d_{yz} $ orbital degeneracy is lifted by the application of hydrostatic pressure. At 60 GPa pressure, in tetragonal pyramid coordination (space group-$ I4mm $) we found reversal of the crystal fields as compared to the tetragonal coordination (space group-$ P4/nmm $) at ambient pressure. Here at 60 GPa pressure $ e_g $ orbitals are energetically higher as compared to the $ t_{2g} $ orbitals (see Fig. \ref{fig:CFS_LiFeAs}C). The same phenomena had been observed in case of a negative charge transfer system \ce{CsAuCl_3} \cite{WF_JPCM}. This metal-semiconductor transition may lead to reversal of the $ p-d $ hybridization contribution to the $ e_g $-$ t_{2g} $ splitting. With the $ t_{2g} $-dominated bands energetically lower than the $ e_g $-dominated bands, the $ p-d $ hybridization may be stronger for $ t_{2g} $ orbitals than $ e_g $ orbitals \cite{WF_JPCM}. At 240 GPa pressure, in hexagonal structure we observe reversal of $ d_{xy} $ and $ d_{x^2-y^2} $ orbitals as compared to CFS at ambient pressure (see Fig. \ref{fig:CFS_LiFeAs}D). Most important splittings between the $ t_{2g} $ and $ e_g $ multiplets are tabulated in the Table-\ref{table:CFS1}. The intra-multiplet splittings are symmetry dependent and can change sign across different symmetries. 

\begin{table}
	\centering
	\begin{tabular}{|M{1.5cm}|P{1.5cm} P{1.5cm} P{1.5cm} P{1.5cm}|}
		\hline
		\multirow{2}{*}{Orbitals}  & \multicolumn{4}{|c|}{Pressure} \\ \cline{2-5} 		
		& 0 GPa & 40 GPa & 60 GPa & 240 GPa \\
		\hline
		$ d_{xz} $   &   &   &   & \\ 
         \hspace{20pt} \large	\}   & 0.0    & 0.130  & 0.072 & 0.147\\	
		$ d_{yz} $   &   &   &   & \\ 	
		\hspace{20pt} \large	\}   & 0.133  & 0.616  &-0.067 & 0.622\\
		$ d_{xy} $   &   &   &   & \\ 	
		\hspace{20pt} \large	\}   & 0.187  & 0.090  & -0.015 & -0.203\\	
   $ d_{x^2-y^2} $   &   &   &   & \\ 
        \hspace{20pt} \large	\}   & 0.111  & 0.574  & -0.066 & 0.698\\	
	   $ d_{z^2} $   &   &   &   & \\ 		
		\hline
	\end{tabular}
	\caption{Crystal field splitting of the orbital levels in LiFeAs at various pressures (in eV).}
	\label{table:CFS1}
\end{table}

\begin{table*}
	\centering
	\begin{tabular}{|M{4cm}|M{3cm}|P{2.5cm}|P{2.5cm}|P{2.5cm}|}
		\hline
		\multirow{2}{*}{\shortstack{Bands considred in \\ Wannier function \\ construction}} 
		& \multirow{2}{*}{Symbol ($ \Delta_{cf} $)} & \multicolumn{3}{|c|}{$ \Delta_{cf} $ in eV at pressure} \\ \cline{3-5} 		
		& &0 GPa & 40 GPa & 240 GPa \\
		\hline
		only effective Fe-$ d $   & $ \epsilon^{(d)}_{t_{2g}}-\epsilon^{(d)}_{e_{g}}$  & 0.331  & 0.831 & 0.610  \\ 
		& & \hspace{1.cm} {\large\}} 0.195 & \hspace{1.cm} {\large\}} 0.516 & \hspace{1.cm} {\large\}} 0.435\\
		Fe-$ d $ and As-$ p $     & $ \epsilon^{(dp)}_{t_{2g}}-\epsilon^{(dp)}_{e_{g}}$  & 0.136  & 0.315  &  0.175 \\
		& & \hspace{1.cm} {\large\}} 0.002 & \hspace{1.cm} {\large\}} 0.014 & \hspace{1.cm} {\large\}} 0.05\\
Fe-$ d $, As-$ p $ and Li-$ s $   & $ \epsilon^{(dps)}_{t_{2g}}-\epsilon^{(dps)}_{e_{g}}$ & 0.134  & 0.301  &  0.125 \\
		\hline
	\end{tabular}
	\caption{Crystal field splitting of the orbital levels in LiFeAs at various pressures.}
	\label{table:CFS2}
\end{table*}

The effect of hybridization between Fe atoms with As ligands as well as purely electrostatic contribution to the crystal potential are included in the onsite energies. We denote onsite energies of Fe-centered Wannier functions by $ \epsilon^{(d)} $, where the superscript indicates that corresponding Wannier functions are obtained using effective Fe-$ d $ orbitals only. Then the onsite energy difference between $ \epsilon^{(d)}_{t_{2g}} $ and $ \epsilon^{(d)}_{e_{g}}$ like Wannier orbitals can be interpreted as the full crystal field splitting $ \Delta_{cf} $. Practically, $ \Delta_{cf} $ is calculated using difference between the average on-site energy of the three $ t_{2g} $ orbitals and that of the two $ e_g $ orbitals \cite{WF_JPCM}. Then we construct second set of Wannier functions, using Fe-$ d $ and As-$ p $ orbitals. We denote the onsite energies of this so called `dp-model' by $ \epsilon^{(dp)} $. Then we construct third set of Wannier functions using Fe-$ d $, As-$ p $ and Li-$ s $ orbitals and denote the onsite energies of this so called `dps-model' by $ \epsilon^{(dps)} $. The splittings on the onsite energies for these three sets are tabulated in Table-\ref{table:CFS2} at atmospheric pressure (0 GPa) and high pressures (40 and 240 GPa). Full crystal field splittings in `d-model' are 0.331 eV, 0.831 eV and 0.631 eV at 0 GPa, 40 GPa and 240 GPa pressure respectively. By a close look one can see that at atmospheric pressure splitting is reduced by 0.195 eV in the second set as compared to the first. This 0.195 eV reduction can be understood as, the contribution to the splitting stemming from hybridization of the central Fe-$ d $ orbitals with $ p $-orbitals of surrounding As ligands. At 40 GPa and 240 GPa this reduction due to $ d-p $ hybridization are 0.516 eV and 0.435 eV respectively (see Table-\ref{table:CFS2}). This increasing contribution of $ d-p $ hybridization into full CFS may be responsible for the lifting of $ d_{xz} $ and $ d_{yz} $ orbital degeneracy. It can be seen that in the `dps-model' (third set) $ t_{2g}-e_g $ energy splitting further reduces by 2 meV, 14 meV and 50 meV at 0 GPa, 40 GPa and 240 GPa pressures respectively as compared to the second set of Wannier orbitals. These contributions are relatively less in the full CFS and can be interpreted as an effect of $ d-s $ hybridization. Therefore, we can conclude that by including such more and more number of bands in the Wannier function construction \emph{i.e.,} by removing the inter-site hybridization effect on different sets of orbitals full CFS can be converged to a value having purely electrostatic contribution.

In summary, full CFS of 0.331 eV in LiFeAs at ambient pressure contains a contribution of 0.195 eV from $ d-p $ hybridization, about 2 meV from $ d-s $ hybridization and the rest $ \sim $ 0.134 eV (40\%) is of purely electrostatic origin. At 40 GPa pressure, full CFS of 0.831 eV contains a contribution of 0.516 eV from $ d-p $ hybridization, about 14 meV from $ d-s $ hybridization and $ \sim $ 0.301 eV (36\%) is of electrostatic origin. At 240 GPa pressure, full CFS of 0.610 eV contains a contribution of 0.435 eV from $ d-p $ hybridization, about 50 meV from $ d-s $ hybridization and the remaining $ \sim $ 0.125 eV (20\%) is of electrostatic origin. This systematic decrease in purely electrostatic contribution in full CFS with increasing pressure, implies larger hybridization effect due to the applied pressure.

\subsubsection*{Crystal field splitting of NaFeAs under pressure}
\begin{figure*}
	\includegraphics[width=0.95\textwidth, height=12cm]{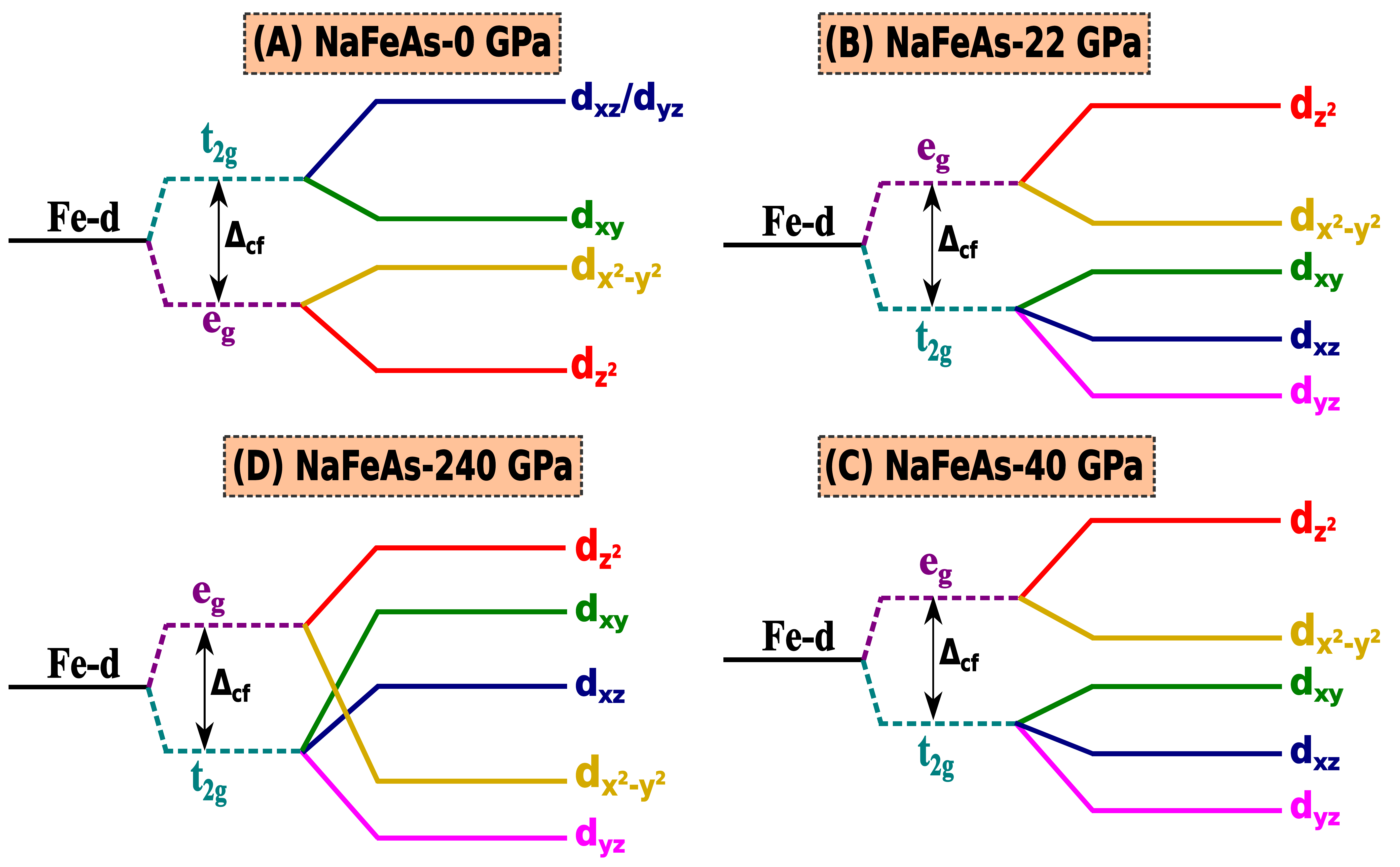}
	\caption{(Color online) Crystal field splittings of NaFeAs at various pressures (cartoon figure, not drawn to scale). Atmospheric pressure: (A) 0 GPa,  high pressures: (B) 22 GPa, (C) 40 GPa, and (D) 240 GPa.}
	\label{fig:CFS_NaFeAs}
\end{figure*}

From Fig. \ref{fig:NaFeAs_PDOS} we find that Fe-$ d $ and As-$ 4p_z $ orbital contributes most in the partial density of states near FL. MLWFs are constructed for each group of bands separately and the resulting MLWFs can be viewed as atomic orbitals arising from hybridization between Fe/As/Na atoms. $ \epsilon^{(d)} $ denote the onsite energies for the Fe-centered WFs corresponding to set of Fe-$ d $ orbitals. The crystal field splitting using this so called `d-model' for NaFeAs at different pressures are presented in Fig. \ref{fig:CFS_NaFeAs}. In tetrahedral coordination, at ambient pressure Fe-$ d $ orbital splits into $ e_{g} (d_{x^2-y^2}, d_{z^2}) $ and $ t_{2g} (d_{xz}, d_{yz}, d_{xy})$ orbitals \cite{de_medici}. Here, $ d_{x^2-y^2}, d_{z^2}, d_{xy} $ orbitals are non degenerate, but $ d_{xz}, d_{yz} $ orbitals are degenerate (see Fig. \ref{fig:CFS_NaFeAs}A). Here we found $ e_g $ orbitals are higher in energy than that of $ t_{2g} $ orbitals. At 22 GPa pressure, in distorted tetrahedral coordination Fe-$ d $ orbital splits into doubly non degenerate $ d_{x^2-y^2}, d_{z^2} $ orbitals and triply non degenerate $ d_{xz}, d_{yz}, d_{xy} $ orbitals (see Fig. \ref{fig:CFS_NaFeAs}B). Here we observed that hydrostatic pressure lifts the orbital degeneracy of $ d_{xz}, d_{yz} $ orbitals completely. Again, this reverse the orbital orientation of $ t_{2g} $ and $ e_g $ orbitals as compared to the tetrahedral coordination ($ e_g $ orbitals are energetically higher as compared to $ t_{2g} $ orbitals). In case of negative charge transfer system \ce{CsAuCl_3}, this similar kind of phenomena has been observed \cite{WF_JPCM}. At 40 GPa pressure there is no change in coordination environment of NaFeAs, hence CFS is least affected (see Fig. \ref{fig:CFS_NaFeAs}C). In octahedral coordination, at 240 GPa pressure $ t_{2g} $ orbitals penetrates in to $ e_g $ orbitals (see Fig. \ref{fig:CFS_NaFeAs}D). Most important splittings between $ t_{2g} $ and $ e_g $ multiplets are tabulated in the Table-\ref{table:CFS3}. The intra-multiplet splittings are symmetry dependent and can change sign across different symmetries.

\begin{table}
	\centering
	\begin{tabular}{|M{1.5cm}|P{1.5cm} P{1.5cm} P{1.5cm} P{1.5cm}|}
		\hline
		\multirow{2}{*}{Orbitals}  & \multicolumn{4}{|c|}{Pressure} \\ \cline{2-5} 		
		& 0 GPa & 22 GPa & 40 GPa & 240 GPa \\
		\hline
		$ d_{xz} $   &   &   &   & \\ 
		\hspace{20pt} \large	\}   & 0.0    & 0.076  & 0.093 & 0.135\\	
		$ d_{yz} $   &   &   &   & \\ 	
		\hspace{20pt} \large	\}   & 0.080  & -0.167  &-0.263 & -0.152\\
		$ d_{xy} $   &   &   &   & \\ 	
		\hspace{20pt} \large	\}   & 0.320  & -0.042  & -0.001 & 0.033\\	
		$ d_{x^2-y^2} $   &   &   &   & \\ 
		\hspace{20pt} \large	\}   & 0.014  & -0.003  & -0.029 & -0.060\\	
		$ d_{z^2} $   &   &   &   & \\ 		
		\hline
	\end{tabular}
	\caption{Crystal field splitting of the orbital levels in NaFeAs at various pressures (in eV).}
	\label{table:CFS3}
\end{table}

\begin{table*}
	\centering
	\begin{tabular}{|M{4cm}|M{3cm}|P{2.5cm}|P{2.5cm}|P{2.5cm}|}
		\hline
		\multirow{2}{*}{\shortstack{Bands considred in \\ Wannier function \\ construction}} 
		& \multirow{2}{*}{Symbol ($ \Delta_{cf} $)} & \multicolumn{3}{|c|}{$ \Delta_{cf} $ in eV at pressure} \\ \cline{3-5} 		
		& &0 GPa & 22 GPa & 40 GPa \\
		\hline
		only effective Fe-$ d $   & $ \epsilon^{(d)}_{t_{2g}}-\epsilon^{(d)}_{e_{g}}$  & 0.380  & 0.129 & 0.160  \\ 
		& & \hspace{1.cm} {\large\}} 0.198 & \hspace{1.cm} {\large\}} 0.114 & \hspace{1.cm} {\large\}} 0.148\\
		Fe-$ d $ and As-$ p $     & $ \epsilon^{(dp)}_{t_{2g}}-\epsilon^{(dp)}_{e_{g}}$  & 0.182  & 0.015  &  0.012 \\
		& & \hspace{1.cm} {\large\}} 0.017 & \hspace{1.cm} {\large\}} 0.007 & \hspace{1.cm} {\large\}} 0.006\\
		Fe-$ d $, As-$ p $ and Na-$ s $   & $ \epsilon^{(dps)}_{t_{2g}}-\epsilon^{(dps)}_{e_{g}}$ &   0.165  & 0.008  &  0.006 \\
		
		\hline
	\end{tabular}
	\caption{Different kinds of hybridization contributions in the crystal field splitting of NaFeAs.}
	\label{table:CFS4}
\end{table*}

Onsite energies include both the effects of hybridization between As/Fe atoms as well as pure electrostatic contribution to the crystal field potential. The onsite energies of Fe-centered Wannier functions are denoted by $ \epsilon^{(d)} $, where the superscript indicates that the corresponding WFs were obtained using effective Fe-$ d $ orbitals as the initial projection. Then using Fe-$ d $ and As-$ p $ orbitals as initial projections, we construct second set of Wannier functions. Here we denote the onsite energies by $ \epsilon^{(dp)} $ and named it as `dp-model'. Then we construct a third set of Wannier functions using Fe-$ d $, As-$ p $ and Na-$ s $ as initial projections and denote the onsite energies by $ \epsilon^{(dps)} $, we named it as `dps-model'. Different hybridization contributions on the onsite energies for these three sets are tabulated in Table-\ref{table:CFS4} at ambient pressure (0 GPa) and at higher pressures (22 and 40 GPa). Full crystal field splitting measured in `d-model' are 0.380 eV, 0.129 eV, and 0.160 eV at 0 GPa, 22 GPa, and 40 GPa pressures respectively. By a more closer look one can see that at ambient pressure splitting is reduced by 0.198 eV in the second set as compared to the first. This 0.198 eV reduction can be understood as the contribution to the splitting, stemming from hybridization of the central Fe-$ d $ orbitals with $ p $-orbitals of surrounding As ligands. At 22 GPa and 40 GPa this reduction due to $ d-p $ hybridization are 0.114 eV and 0.148 eV respectively. Hence, the contribution of $ d-p $ hybridization into CFS increases from $ \sim $52\% to 88\% with increasing pressure and increases further to 92.5\% at 40 GPa. This increasing $ d-p $ hybridization contribution into full CFS may be responsible for $ d_{xz} $ and $ d_{yz} $ orbital degeneracy lifting. It can be seen that in the third set (`dps-model') $ t_{2g}-e_g $ energy splitting further reduces by 17 meV, 8 meV, and 6 meV at 0 GPa, 22 GPa, and 40 GPa pressures respectively as compared to `dp-model'. In the full CFS these contributions are relatively less and can be interpreted as an effect of $ d-s $ hybridization. Therefore, we can conclude that by removing the inter-site hybridization effect on different sets of orbitals \emph{i.e.,} including such more and more number of bands in the WF construction full CFS can be converged to a value having purely electrostatic contribution.

In summary, full CFS of 380 meV in NaFeAs at ambient pressure contains the contributions of 198 meV from $ d-p $ hybridization, about 17 meV from $ d-s $ hybridization and the rest $ \sim $ 165 meV ($ \sim $43\%) is of purely electrostatic origin. At 22 GPa pressure, full CFS of 129 meV contains the contributions of 114 meV from $ d-p $ hybridization, about 8 meV from $ d-s $ hybridization and 8 meV ($ \sim $ 6\%) is of electrostatic origin. At 40 GPa pressure, full CFS of 160 meV contains the contributions of 148 meV from $ d-p $ hybridization, about 6 meV from $ d-s $ hybridization and remaining 6 meV ($ \sim $ 3.7\%) is of electrostatic origin. This systematic decrease in purely electrostatic contribution in full CFS with increasing pressure, implies larger hybridization effect due to the applied pressure.

\subsubsection*{Visualization of Wannier functions}
\begin{figure*}
	\centering
	\includegraphics[width=14cm, height=14cm]{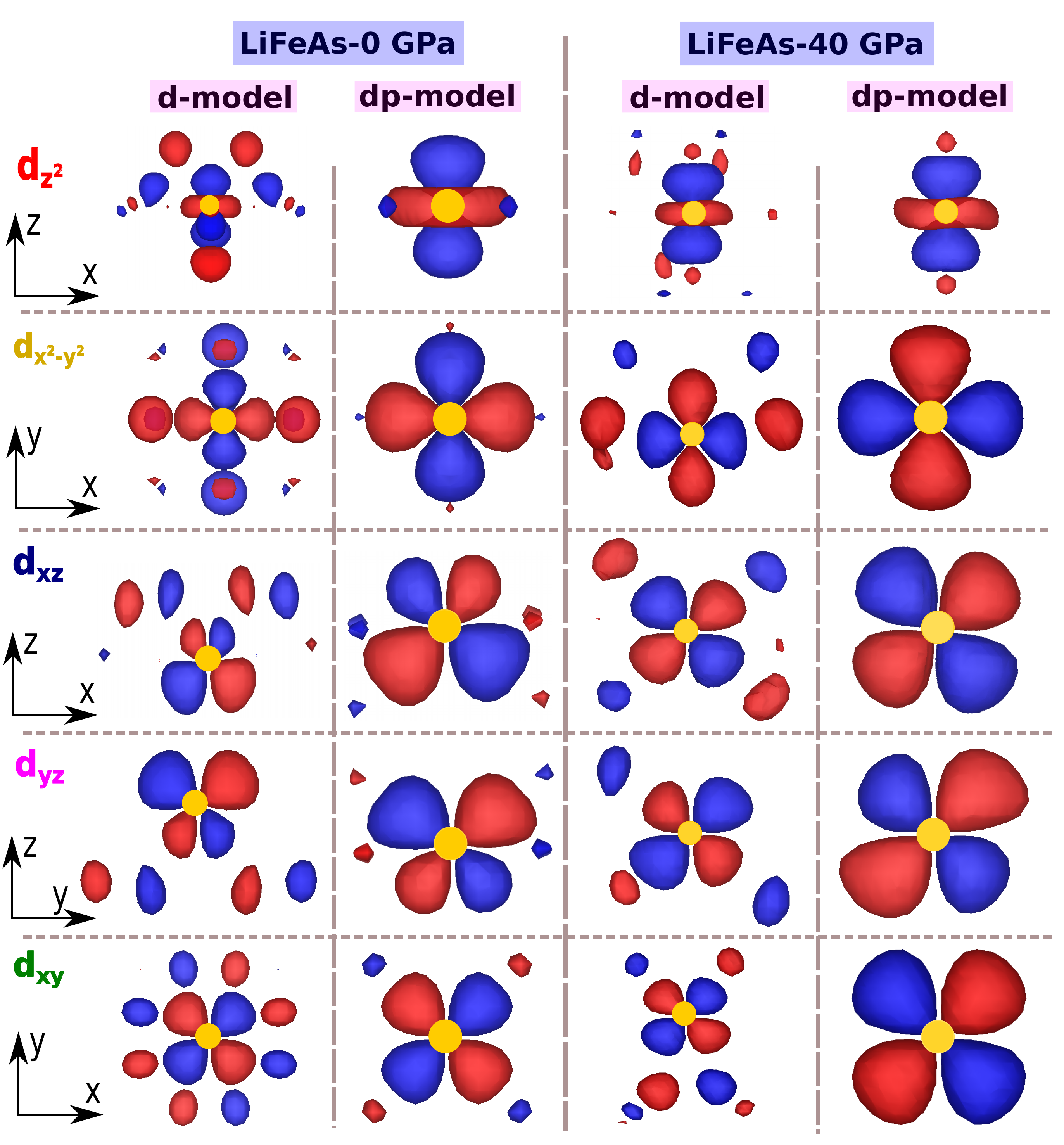}
	\caption{(Color online) Representative members from two different sets of Maximally localized Wannier functions of LiFeAs at various pressures. In d-model Fe-$ d $ orbitals are considered for MLWF construction; whereas in dp-model Fe-$ d $ and As-$ p $ orbitals are considered for MLWF construction. Yellow dot represent Fe atom, center of the MLWFs. The spacial orientation is indicated in all figures. VESTA \cite{vesta} package is used for the visualization.}
	\label{fig:wannier_Li}
\end{figure*}

The resulting WFs for LiFeAs in the `d-model' and `dp-model' at ambient pressure and 40 GPa pressure are shown in Fig. \ref{fig:wannier_Li}. All the WFs are centered on Fe-$ d $ sites. In `dp-model' Fe-$ d $ like WFs contain only minimal contributions for As-$ p $ orbitals situated at the surrounding ligands. Overall, in `dp-model' WFs are much more similar to the atomic orbitals as compared to the WFs of `d-model'.

\begin{figure*}
	\centering
	\includegraphics[width=14cm, height=15cm]{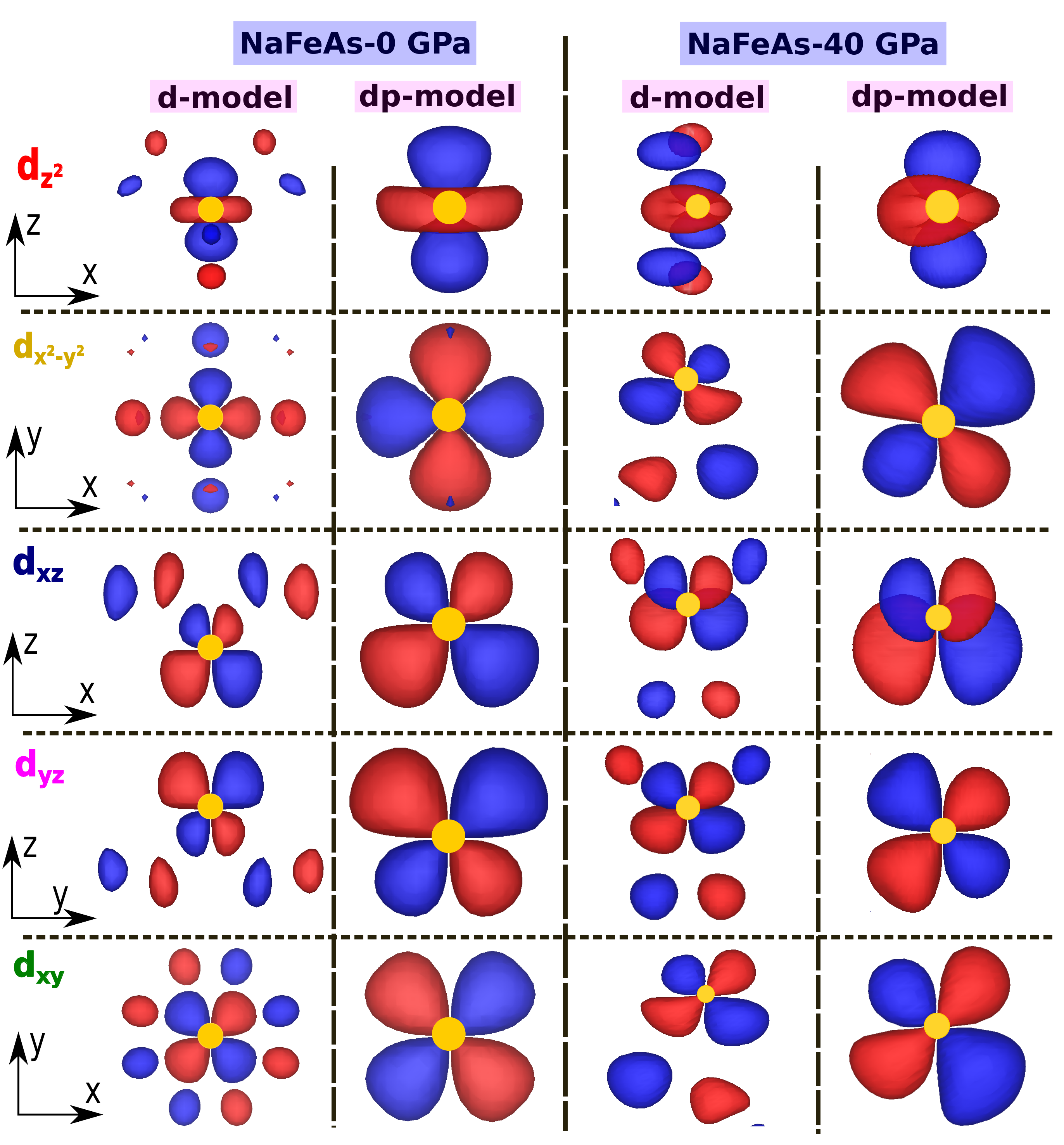}
	\caption{(Color online) Representative members from two different sets of Maximally localized Wannier functions of NaFeAs at various pressures. In d-model Fe-$ d $ orbitals are considered for MLWF construction; whereas in dp-model Fe-$ d $ and As-$ p $ orbitals are considered for MLWF construction. Yellow dot represent Fe atom, center of the MLWFs. The spacial orientations are indicated in all figures. VESTA \cite{vesta} package is used for the visualization.}
	\label{fig:wannier_Na}
\end{figure*}

The resulting WFs for NaFeAs in the `d-model' and `dp-model' are shown in Fig. \ref{fig:wannier_Na}, at ambient pressure and 40 GPa pressure. Like LiFeAs, here all the WFs are centered on Fe-$ d $ sites. In `dp-model' Fe-$ d $ like Wannier functions contain only minimal contributions from As-$ p $ orbitals situated at the surrounding ligands. Overall, in the `dp-model' WFs are much more similar to the atomic orbitals as compared to the Wannier functions of `d-model', following the same trend as in LiFeAs. In the `dp-model', at 40 GPa pressure WFs are more distorted as compared to that at ambient pressure due to the distorted tetrahedral coordination of the central Fe atom. We must note that, the chosen iso-surface value used for WF visualization decides the visible size of admixtures in Figs. \ref{fig:wannier_Li}, \ref{fig:wannier_Na}.

\subsection{Low energy tight binding model of LiFeAs} \label{tb_Li}
\begin{figure}[H]
		\centering
		\includegraphics[width=8cm, height=5.5cm]{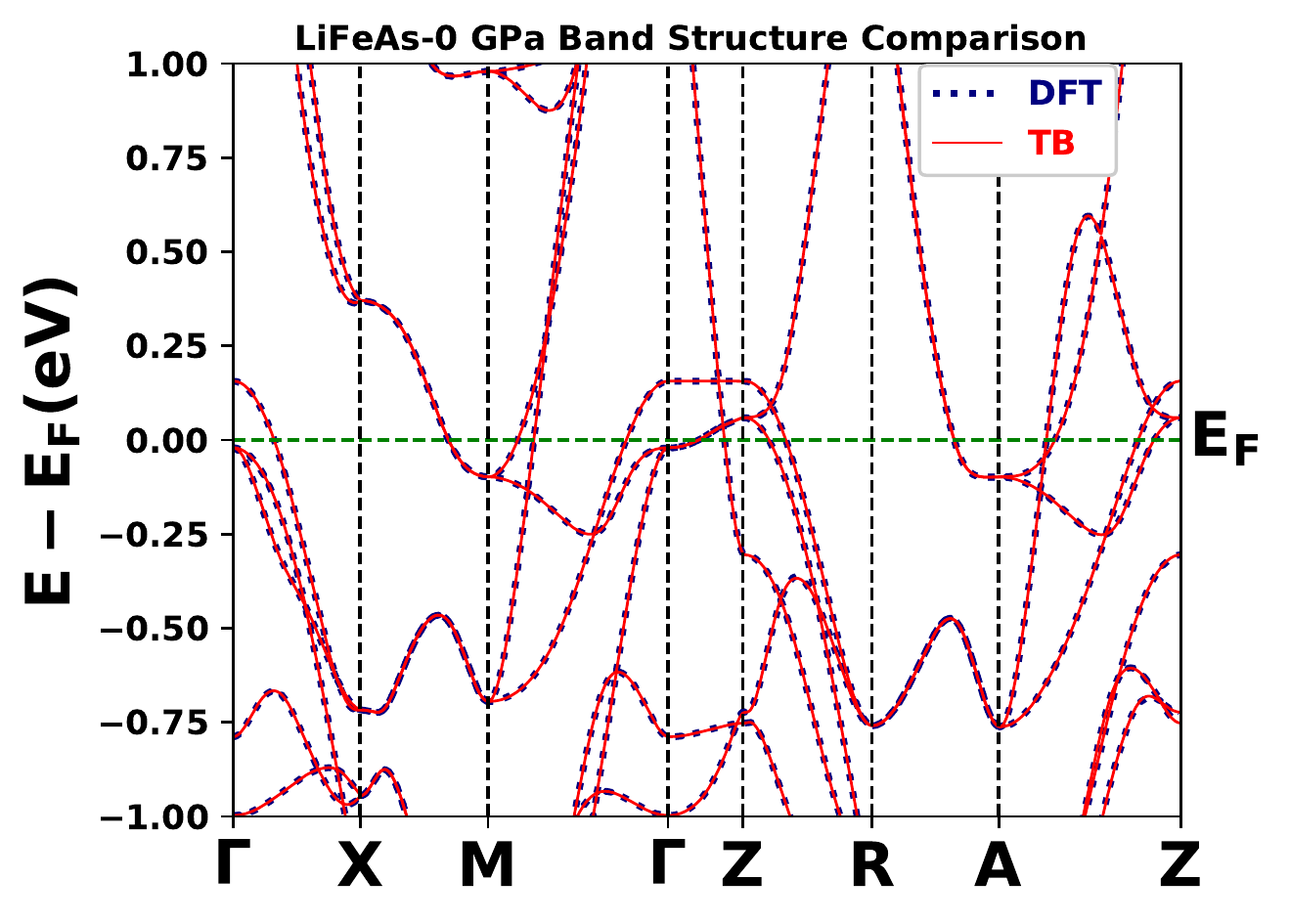}
		\caption{(Color online) Tight binding fitted band structure of LiFeAs at ambient pressure. Fermi energy level is set to zero.}
		\label{fig:LiFeAs-0GPa_BS_DFT+TB}
	\end{figure}

\begin{figure}[H]
		\centering
		\includegraphics[width=8cm, height=5.5cm]{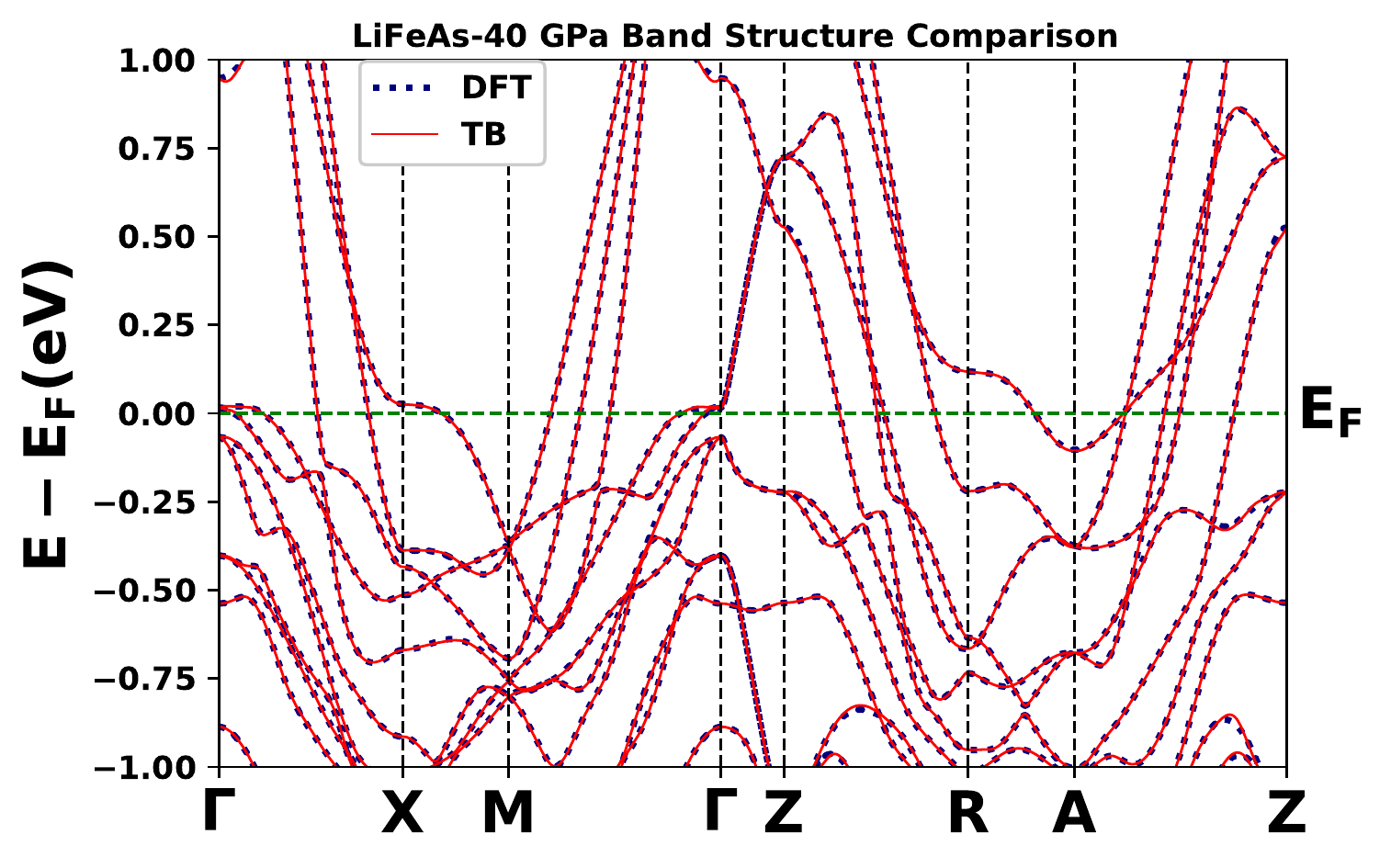}
		\caption{(Color online) Tight binding fitted band structure of LiFeAs at 40 GPa pressure. Fermi energy level is set to zero.}
		\label{fig:LiFeAs-40GPa_BS_DFT+TB}
	\end{figure}

\begin{figure}[H]
		\centering
		\includegraphics[width=8cm, height=5.5cm]{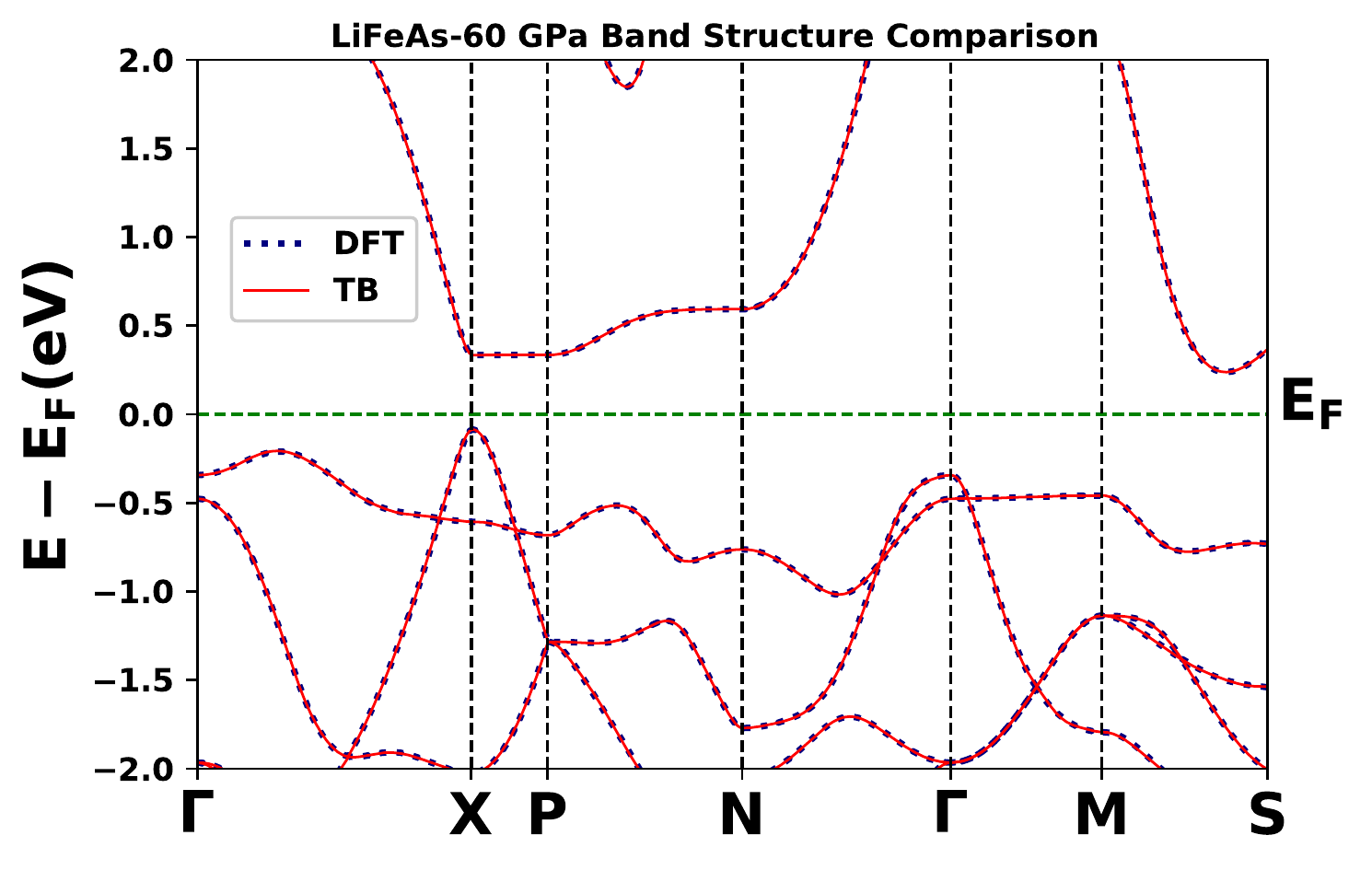}
		\caption{(Color online) Tight binding fitted band structure of LiFeAs at 60 GPa pressure. Fermi energy level is set to zero.}
		\label{fig:LiFeAs-60GPa_BS_DFT+TB}
	\end{figure}

\begin{figure}
		\centering
		\includegraphics[width=9cm, height=5.5cm]{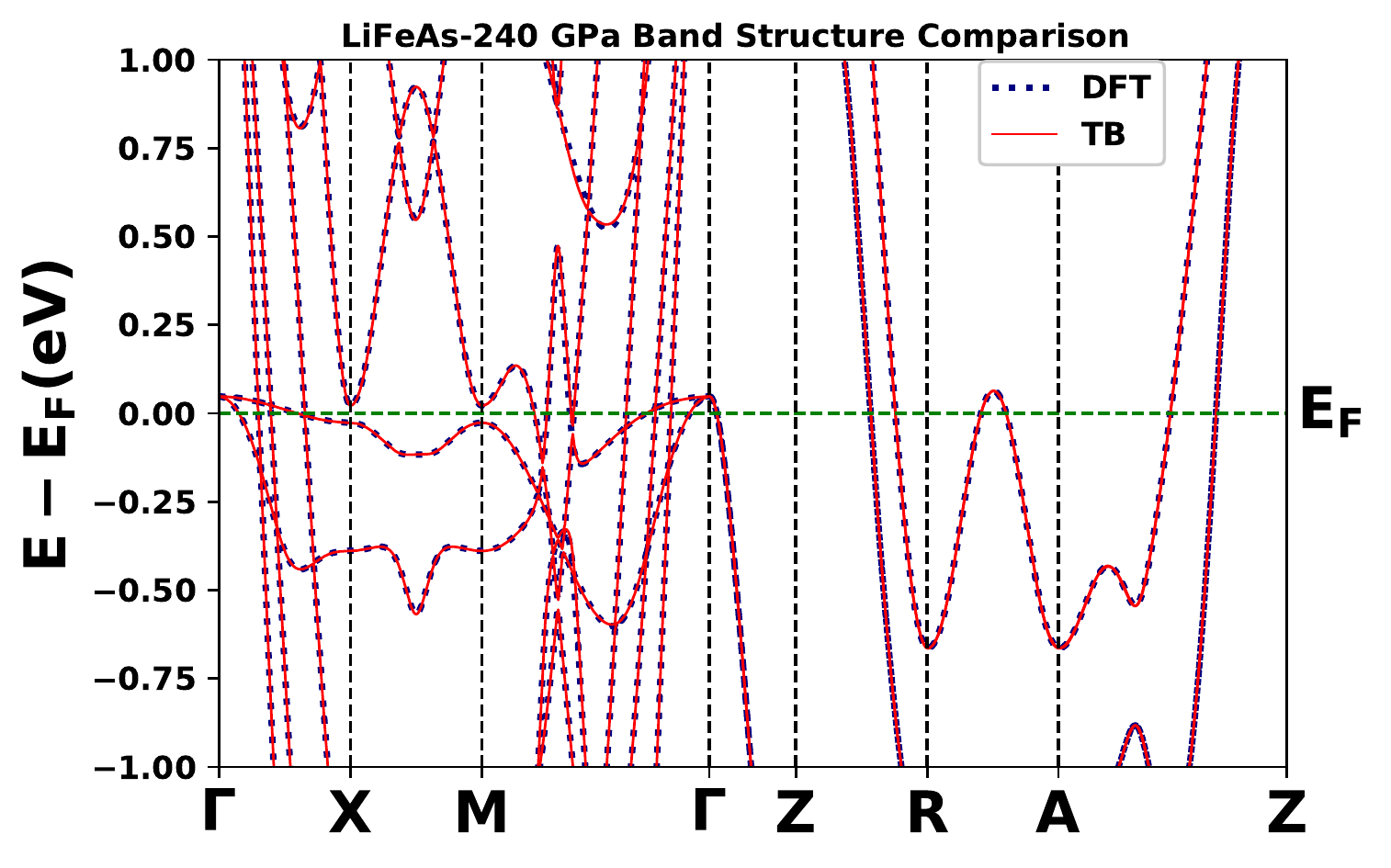}
		\caption{(Color online) Tight binding fitted band structure of LiFeAs at 240 GPa pressure. Fermi energy level is set to zero.}
		\label{fig:LiFeAs-240GPa_BS_DFT+TB}
	\end{figure}

As we have mentioned earlier (section-\ref{sec_cfs}), the matrix elements in eq. (\ref{eq:hopping}) can be interpreted as the hoping amplitude between the two Wannier orbitals centered at different atomic sites. We obtain these hopping elements by constructing the maximally localized Wannier functions using effective Fe-$ d $ orbitals in the finite energy window above and below the FL (-1.0 eV to 1.0 eV) except at 60 GPa pressure. At 60 GPa, we have considered both the Fe-$ d $ and As-$ p $ orbitals for the well converged MLWF construction. Effectively larger presence of As-$ p $ orbitals PDOS (see Fig.-\ref{fig:LiFeAs_PDOS}C) at valance state may be responsible for this effect. The tight binding fitted band structures (red dashed curves) is compared with the DFT derived ones (blue continuous curves) in the Figs.  \ref{fig:LiFeAs-0GPa_BS_DFT+TB}-\ref{fig:LiFeAs-240GPa_BS_DFT+TB} at different pressures. Here we obtain a perfect fitting for low energy bands in all the studied cases. We have presented the nearest-neighbours (NN) and the next-nearest-neighbours (NNN) intra as well as inter orbital hopping amplitudes in Table-\ref{table:hopping_Li} for the different crystallographic symmetries. These hopping amplitudes are extracted from tight binding fitted band structures shown in the Figs. \ref{fig:LiFeAs-0GPa_BS_DFT+TB}-\ref{fig:LiFeAs-240GPa_BS_DFT+TB} and can be used to build effective low energy simplified model Hamiltonians.



\begin{table*}
	\centering
	\begin{tabular}{|M{1.3cm}|P{1.2cm}|P{1.3cm} P{1.3cm} P{1.3cm} P{1.3cm} P{1.3cm}|P{1.3cm} P{1.3cm} P{1.3cm} P{1.3cm} P{1.3cm}|}
		\hline
		\multirow{2}{*}{Pressure} &\multirow{2}{*}{orbitals} 
		& \multicolumn{5}{|c|}{nearest-neighbours (NN) hopping amplitudes} 
		& \multicolumn{5}{|c|}{ next-nearest-neighbours (NNN) hopping amplitudes}\\ \cline{3-12} 		
		& & $d_{z^2}$ & $d_{xz}$ & $d_{yz}$ & $d_{x^2-y^2}$ & $d_{xy}$ & $d_{z^2}$ & $d_{xz}$ & $d_{yz}$ & $d_{x^2-y^2}$ & $d_{xy}$\\
		\hline
		0 GPa & $d_{z^2}$ & -29.04 & 0.0 & -190.01 & 114.32 & 0.0 & -22.35 & -11.09 & 11.09 & 0.0& -38.77\\ 
		      & $d_{xz}$ & 0.0 & 125.88 & 0.0 & 0.0 & 151.40 & 11.09 & -3.14 & 0.0 & 2.0 & 2.59 \\
		      & $d_{yz}$ & 190.01 & 0.0 & 385.03 & 88.77 & 0.0 & -11.09 & 0.0 & -3.14 & 2.0 & -2.59 \\
		 & $d_{x^2-y^2}$ & 114.32 & 0.0 & -88.77 & 131.01 & 0.0 & 0.0 & -2.00 & -2.00 & 2.19 & 0.0 \\
		      & $d_{xy}$ & 0.0 & -151.40 & 0.0 & 0.0 & -60.67 & -38.77 & -2.59 & 2.59 & 0.0 & -38.42 \\
		\hline
		40 GPa & $d_{z^2}$ & -165.36 & 432.89 & 381.86 & 429.07 & 538.53 & -0.86 & -4.24 & 1.79 & 21.04 & -1.17 \\
		& $d_{xz}$ & 36.69 & 258.05 & 179.41 & 83.94 & 187.82 & -0.61 & 0.41 & 2.39 & -0.62 & -1.31 \\
		& $d_{yz}$ & -39.37 & 159.86 & -0.44 & 162.73 & 59.14 & -1.49 & -0.44 & 1.37 & 3.85 & 0.51 \\
		& $d_{x^2-y^2}$ & -38.71 & -0.51 & -33.94 & 2.48 & -8.62 & 1.08 & -0.12 & -0.58 & 0.32 & 0.18 \\
		& $d_{xy}$ & -62.77 & -46.44 & -6.96 & -13.64 & 36.35 & -0.56 & 2.08 & -1.01 & 1.11 & 0.68 \\
		\hline
		60 GPa & $d_{z^2}$ & 201.52 & 81.64 & -5.58 & 14.20 & -1.82 & 0.27 & -28.57 & 0.01 & 2.48 & -2.91 \\ 
		& $d_{xz}$ & -76.64 & -219.09 & 44.23 & -53.50 & 44.75 & 36.74 & -12.41 & -0.08 & -36.60 & 10.36 \\
		& $d_{yz}$ & -5.13 & -44.38 & 954.84 & -252.08 & 1431.76 & 6.91 & 0.85 & -56.99 & 0.12 & -1.98 \\
		& $d_{x^2-y^2}$ & -13.63 & -53.71 & 251.95 & -456.62 & 255.76 & -277.44 & -1086.84 & -1.42 & 386.08 & 207.24 \\
		& $d_{xy}$ & -1.37 & -44.70 & 1431.63 & -255.86 & 997.0 & -932.85 & -105.94 & -1.75 & -14.44 & 197.41 \\
		\hline
		240 GPa & $d_{z^2}$ & 88.82 & -41.50 & 21.20 & 207.92 & 106.82 & -23.36 & -39.23 & -168.31 & -50.37 & 215.81\\ 
		& $d_{xz}$ & 23.80 & 60.07 & -4.07 & -204.98 & 164.87 & -7.21 & 27.82 & -12.95 & 16.10 & -8.07 \\
		& $d_{yz}$ & -590.88 & 1296.80 & 320.87 & 449.46 & 11.36 & 52.21 & 51.04 & 37.28 & -7.35 & -302.78 \\
		& $d_{x^2-y^2}$ & 58.22 & -93.91 & 14.73 & 520.71 & 36.27 & -17.78 & -20.40 & 16.02 & 25.96 & 9.90 \\
		& $d_{xy}$ & 193.38 & 67.89 & -8.50 & -334.54 & 448.27 & -4.61 & -0.63 & -26.10 & 3.71 & -43.45 \\
		\hline
	\end{tabular}
	\caption{Nearest and next-nearest neighbour hopping amplitudes of LiFeAs at various pressures (in meV).}
	\label{table:hopping_Li}
\end{table*}

From Table-\ref{table:hopping_Li} it is evident that at atmospheric pressure NN intra-orbital hopping is maximum in the $ d_{yz} $ orbitals ($ \sim $ 385 meV), while inter-orbital hopping is maximum between the $ d_{xz} $ and $ d_{z^2} $ orbitals ($ \sim $ 190 meV). In case of NNN interaction intra-orbital hopping is maximum in the $ d_{xy} $ orbitals (sign neglected) and inter-orbital hopping is maximum between the $ d_{xy} $ and $ d_{z^2} $ orbitals ($ \sim $ 38.77 meV). At 40 GPa pressure NN intra-orbital hopping is maximum in the $ d_{xz} $ orbitals ($ \sim $ 258 meV), while inter-orbital hopping is maximum between the $ d_{xy} $ and $ d_{z^2} $ orbitals ($ \sim $ 538 meV). In case of NNN interaction intra-orbital hopping amplitudes are very less as compared to the NN hopping amplitudes. At 60 GPa, NN intra-orbital hopping is maximum in the $ d_{xy} $ orbitals ($ \sim $ 997 meV), while inter-orbital hopping is maximum between the $ d_{xy} $ and $ d_{yz} $ orbitals ($ \sim $ 1431 meV). In case of NNN interaction intra-orbital hopping is maximum in the $ d_{x^2-y^2} $ orbitals ($ \sim $ 386 meV) and inter-orbital hopping is maximum between the $ d_{xz} $ and $ d_{x^2-y^2} $ orbitals (sign neglected). At very high pressure around 240 GPa, NN intra-orbital hopping is maximum in the $ d_{x^2-y^2} $ orbitals ($ \sim $ 520 meV), while inter-orbital hopping is maximum between the $ d_{xz} $ and $ d_{yz} $ orbitals ($ \sim $ 1297 meV). In case of NNN interaction intra-orbital hopping is maximum in the $ d_{xy} $ orbitals and inter-orbital hopping is maximum between the $ d_{xy} $ and $ d_{yz} $ orbitals. Therefore, the main features observed in intra as well as inter-orbital hopping amplitudes may be summarized as-(i) at ambient pressure intra-orbital hopping amplitudes are dominant, while inter-orbital hopping amplitudes becomes predominant with the increasing pressure, (ii) intra-orbital as well as inter-orbital hopping amplitudes are maximum in the semi-conducting phase (at 60 GPa), (iii) in metallic phase intra/inter-orbital hopping amplitudes increases with the increasing pressure, (iv) hopping amplitudes are strongly dependent on crystallographic symmetry.


\subsection{Low energy tight binding model of NaFeAs under pressure} \label{tb_Na}
\begin{figure}
	\centering
	\includegraphics[width=8cm, height=5.5cm]{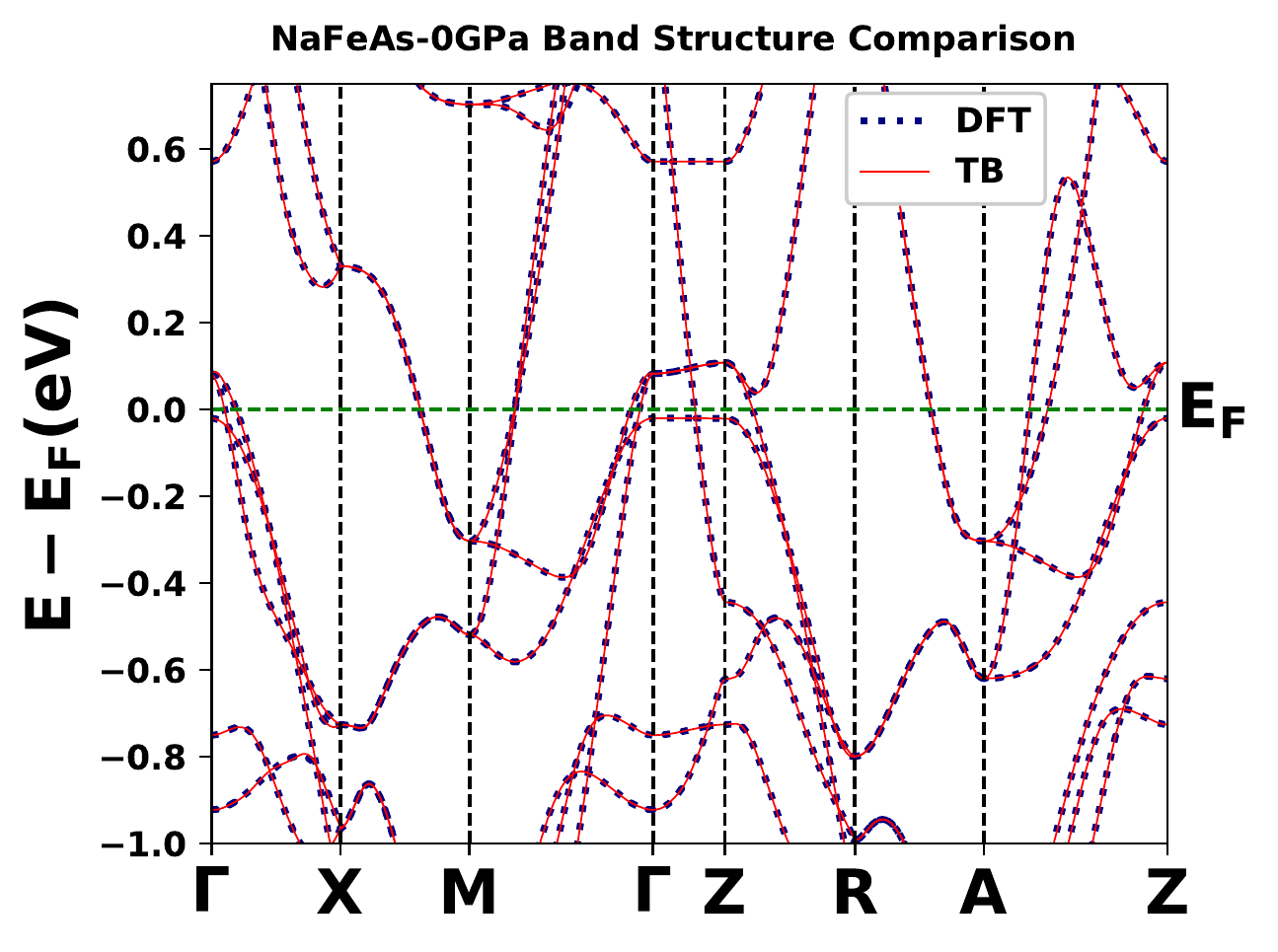}
	\caption{(Color online) Tight binding fitted band structure of NaFeAs at ambient pressure. Fermi energy level is set to zero.}
	\label{fig:NaFeAs-0GPa_BS_DFT+TB}
\end{figure}
\begin{figure}
	\centering
	\includegraphics[width=8cm, height=5.5cm]{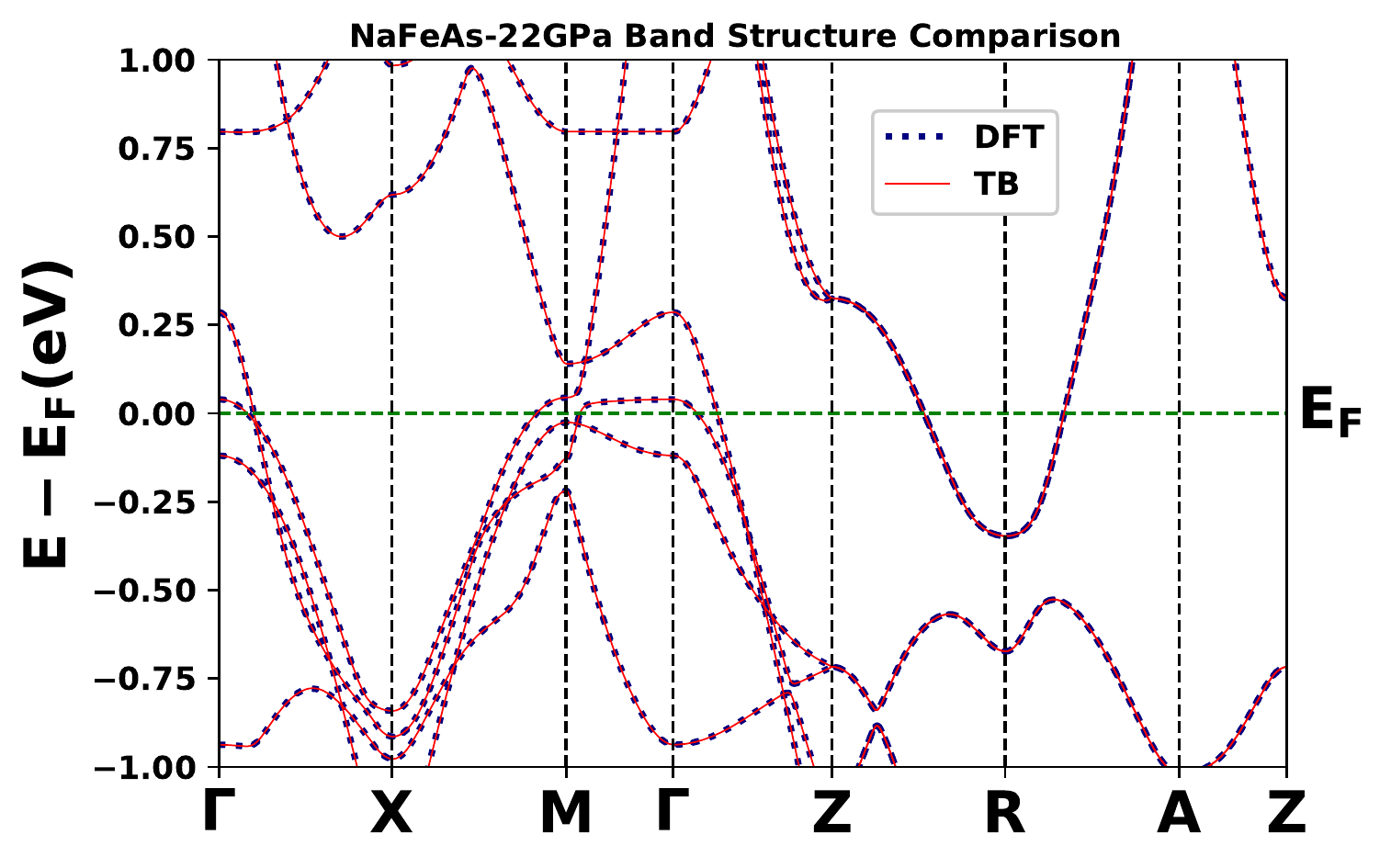}
	\caption{(Color online) Tight binding fitted band structure of NaFeAs at 22 GPa pressure. Fermi energy level is set to zero.}
	\label{fig:NAFeAs-22GPa_BS_DFT+TB}
\end{figure}

\begin{figure}
	\centering
	\includegraphics[width=8cm, height=5.5cm]{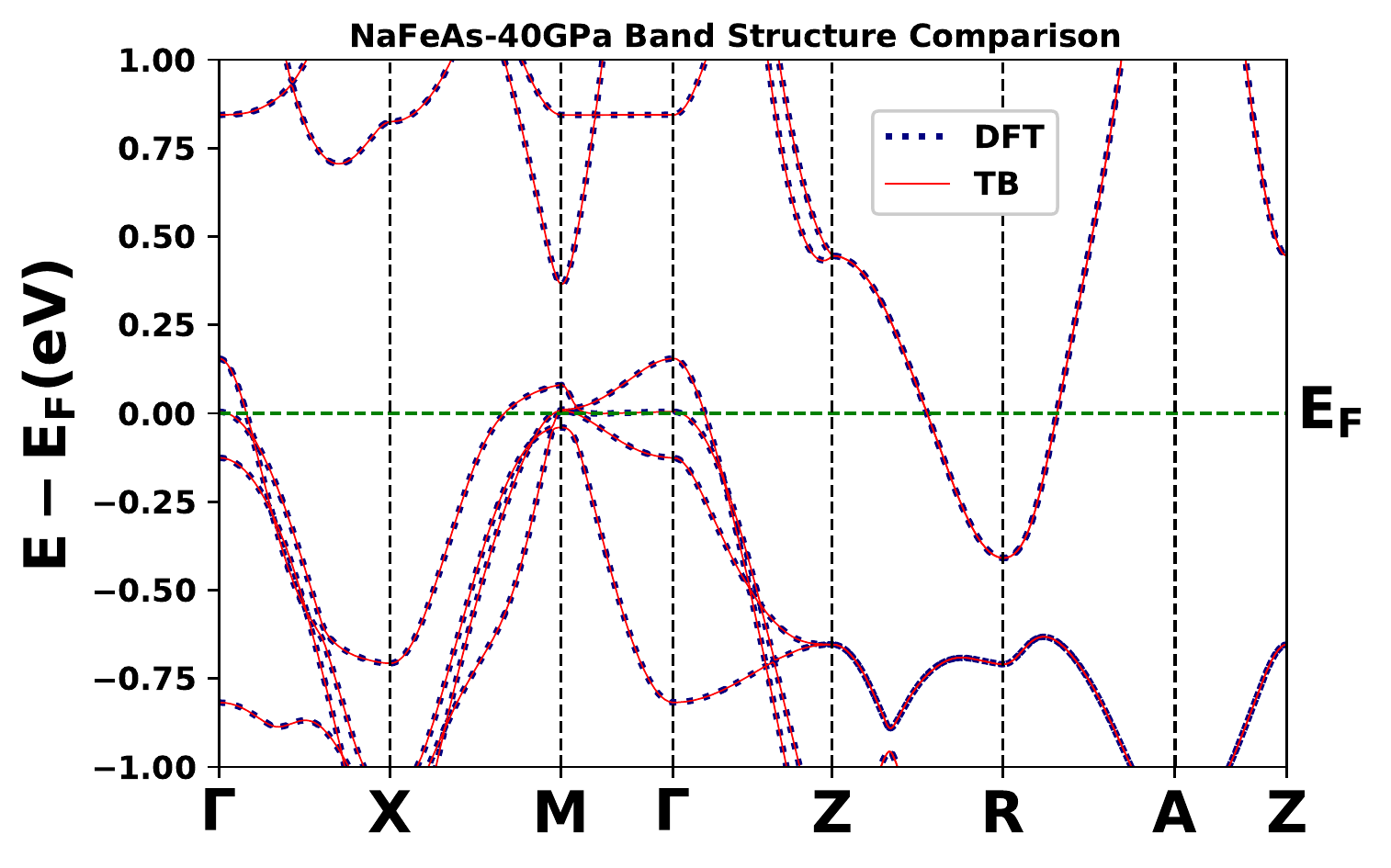}
	\caption{(Color online) Tight binding fitted band structure of NaFeAs at 40 GPa pressure. Fermi energy level is set to zero.}
	\label{fig:NaFeAs-40GPa_BS_DFT+TB}
\end{figure}
\begin{figure}
	\centering
	\includegraphics[width=8cm, height=5.5cm]{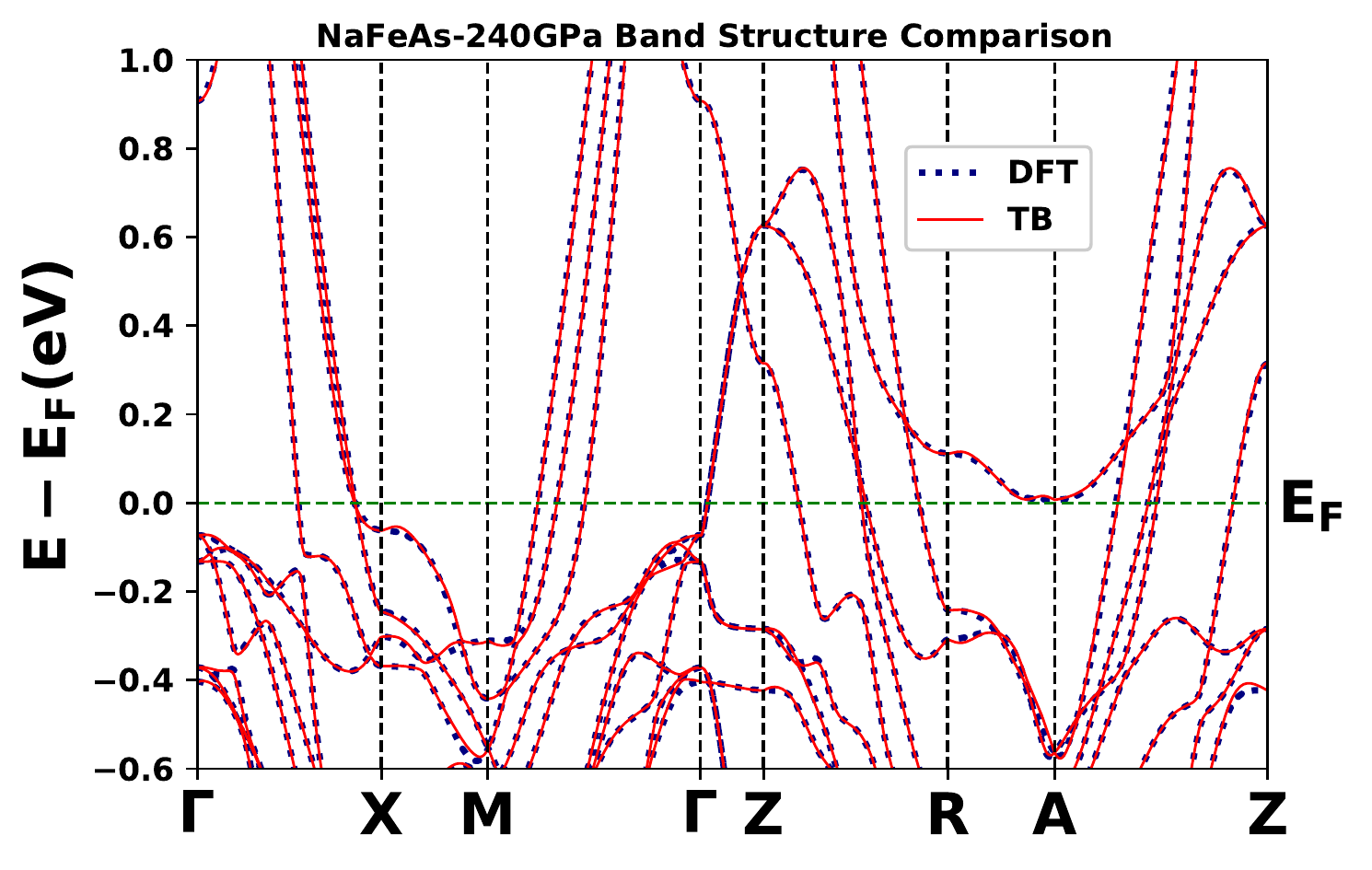}
	\caption{(Color online) Tight binding fitted band structure of NaFeAs at 240 GPa pressure. Fermi energy level is set to zero.}
	\label{fig:NaFeAs-240GPa_BS_DFT+TB}
\end{figure}

For NaFeAs, MLWFs constructed using effective Fe-$ d $ orbitals as initial projections in the finite energy window above and below the FL (-1.0 eV to 1.0 eV) are used to obtain the hopping elements. DFT derived low energy band structures (blue dotted curves) are compared with the tight binding fitted ones (red dashed curve) in Figs.  \ref{fig:NaFeAs-0GPa_BS_DFT+TB}-\ref{fig:NaFeAs-240GPa_BS_DFT+TB} at different pressures. In all the studied cases we find almost perfect fitting near the FL. We have presented the nearest-neighbours (NN) and next-nearest-neighbours (NNN) hopping amplitudes (intra as well as inter orbital) in Table-\ref{table:hopping_Na} for the different crystallographic symmetries. These hopping amplitudes are extracted from the tight binding fitted band structures shown in Figs. \ref{fig:NaFeAs-0GPa_BS_DFT+TB}-\ref{fig:NaFeAs-240GPa_BS_DFT+TB}.

\begin{table*}
	\centering
	\begin{tabular}{|M{1.3cm}|P{1.2cm}|P{1.3cm} P{1.3cm} P{1.3cm} P{1.3cm} P{1.3cm}|P{1.3cm} P{1.3cm} P{1.3cm} P{1.3cm} P{1.3cm}|}
		\hline
		\multirow{2}{*}{Pressure} &\multirow{2}{*}{Orbitals} 
		& \multicolumn{5}{|c|}{Nearest-neighbours (NN) hopping amplitudes} 
		& \multicolumn{5}{|c|}{ Next-nearest-neighbours (NNN) hopping amplitudes}\\ \cline{3-12} 		
		& & $d_{z^2}$ & $d_{xz}$ & $d_{yz}$ & $d_{x^2-y^2}$ & $d_{xy}$ & $d_{z^2}$ & $d_{xz}$ & $d_{yz}$ & $d_{x^2-y^2}$ & $d_{xy}$\\
		\hline
		0 GPa & $d_{z^2}$ & -25.55 & -4.40 & -154.53 & 112.88 & -0.27 & -8.07 & -16.95 & -0.40 & 12.95 & -5.72\\ 
		& $d_{xz}$ & 4.25 & 116.50 & 6.02 & 1.29 & 124.67 & -16.58 & -31.48 & 18.33 & 13.65 & 32.83 \\
		& $d_{yz}$ & 160.06 & 7.07 & 368.36 & 61.06 & -3.54 & -0.69 & 19.20 & 14.28 & 4.38 & -14.52 \\
		& $d_{x^2-y^2}$ & 111.35 & -1.25 & -62.97 & 77.30 & -0.26 & 12.88 & 14.17 & 3.86 & 35.53 & -18.79 \\
		& $d_{xy}$ & 0.26 & -122.03 & 4.03 & 0.21 & -55.72 & -5.89 & 32.86 & 15.96 & -19.07 & 8.82 \\
		\hline
		22 GPa & $d_{z^2}$ & -43.53 & 229.56 & -148.38 & -217.34 & 68.34 & -48.75 & -94.83 & 260.30 & 53.83 & -224.39 \\
		& $d_{xz}$ & 229.56 & 26.27 & 25.37 & 40.32 & -43.41 & -94.83 & -640.44 & 187.50 & 499.95 & 3.66 \\
		& $d_{yz}$ & -148.38 & 25.36 & -692.25 & 28.85 & 524.98 & 260.30 & 187.50 & -25.19 & -160.06 & 41.52 \\
		& $d_{x^2-y^2}$ & -217.35 & 40.32 & 28.85 & -164.68 & 96.38 & 53.83 & 499.95 & -160.06 & -88.51 & 88.07 \\
		& $d_{xy}$ & 68.34 & -43.41 & 524.98 & 96.38 & -102.25 & -224.39 & 3.66 & 41.52 & 88.07 & -172.97 \\
		\hline
		40 GPa & $d_{z^2}$ & -52.75 & 239.70 & -169.14 & -250.89 & 7.93 & -52.93 & -78.82 & 283.63 & 6.26 & -251.30 \\ 
		& $d_{xz}$ & 239.70 & 49.43 & 0.58 & 18.54 & -45.90 & -78.82 & -663.13 & 258.05 & 528.23 & 8.62 \\
		& $d_{yz}$ & -169.14 & 0.58 & -756.46 & 5.96 & 578.97 & 283.63 & 258.05 & -43.55 & -238.72 & 17.08 \\
		& $d_{x^2-y^2}$ & -250.88 & 18.54 & 5.96 & -186.94 & 102.05 & 6.26 & 528.23 & -238.72 & -103.96 & 101.19 \\
		& $d_{xy}$ & 7.93 & -45.90 & 578.97 & 102.05 & -104.84 & -251.30 & 8.62 & 17.08 & 101.18 & -187.19 \\
		\hline
		240 GPa & $d_{z^2}$ & 148.40 & -3.14 & 132.12 & 659.91 & -6.25 & 140.68 & 209.68 & -112.94 & -302.82 & 496.37 \\ 
		& $d_{xz}$ & 4.30 & 85.06 & 0.82 & 2.72 & -204.98 & 123.09 & -292.60 & 217.04 & -200.45 & 154.97 \\
		& $d_{yz}$ & 83.34 & -0.85 & -471.02 & 188.67 & -2.84 & -76.95 & 196.52 & -35.21 & -43.33 & -299.33 \\
		& $d_{x^2-y^2}$ & 300.69 & 2.19 & 335.64 & -372.37 & 12.25 & -146.28 & -107.01 & 157.91 & -2.77 & 214.04 \\
		& $d_{xy}$ & -2.74 & 90.24 & -0.19 & 1.62 & 135.88 & 237.85 & 297.83 & -119.16 & 235.90 & -262.53 \\
		\hline
	\end{tabular}
	\caption{Nearest and next-nearest neighbour hopping amplitudes of NaFeAs at various pressures (in meV).}
	\label{table:hopping_Na}
\end{table*}

From Table-\ref{table:hopping_Na} it is evident that at ambient pressure NN intra-orbital hopping is maximum in between the $ d_{yz} $ orbitals ($ \sim $ 368 meV), while inter-orbital hopping is maximum between $ d_{yz} $ and $ d_{z^2} $ orbitals ($ \sim $ 160 meV). In case of NNN interaction intra-orbital hopping is maximum in the $ d_{x^2-y^2} $ orbitals and inter-orbital hopping is maximum between $ d_{xy} $ and $ d_{xz} $ orbitals ($ \sim $ 32.86 meV). At 40 GPa, NN intra-orbital hopping is maximum in $ d_{yz} $ orbitals ($ \sim $ 692 meV, sign neglected), while inter-orbital hopping is maximum between $ d_{xy} $ and $ d_{yz} $ orbitals ($ \sim $ 525 meV). In case of NNN interaction intra-orbital hopping is maximum in between the $ d_{xz} $ orbitals ($ \sim $ 640 meV) and inter-orbital hopping is maximum between $ d_{xz} $ and $ d_{x^2-y^2} $ orbitals. At 40 GPa pressure, NN intra-orbital hopping is maximum in between $ d_{yz} $ orbitals ($ \sim $ 756 meV, sign neglected), while inter-orbital hopping is maximum between $ d_{xy} $ and $ d_{yz} $ orbitals ($ \sim $ 579 meV). In case of NNN interaction intra-orbital hopping amplitudes are also comparable with NN hopping amplitudes. At relatively high pressure around 240 GPa, NN intra-orbital hopping is maximum in between $ d_{yz} $ orbitals ($ \sim $ 471 meV), while inter-orbital hopping is maximum between $ d_{x^2-y^2} $ and $ d_{yz} $ orbitals ($ \sim $ 660 meV). In case of NNN interaction intra-orbital hopping is maximum in between $ d_{xz} $ orbitals and inter-orbital hopping is maximum between $ d_{xy} $ and $ d_{z^2} $ orbitals. 

Therefore, the observed features in intra as well as inter-orbitals hopping amplitudes can be summarized as-(i) at relatively lower pressure (upto 40 GPa) intra-orbital hopping amplitudes are larger, while inter-orbital hopping amplitudes becomes predominant at very high pressure, (ii) in metallic phase intra/inter-orbital hopping amplitudes increases with increasing pressure from 22 to 40 GPa, (iii) intra-orbital as well as inter-orbital hopping amplitudes are maximum at 40 GPa, (iv) hopping amplitudes are strongly crystallographic symmetry dependent, hence directly related to local structural environment of Fe atom along with its hybridization with surrounding As ligands. Local structural environment of Fe atom at various pressures can be understood from Fe $K$-edge core electron spectroscopic study of Ref. \cite{soumya3}.

\section{Conclusions} \label{conclusion}
We have presented detailed electronic structure study using DFT based first principles calculations for the iron based superconducting materials LiFeAs and NaFeAs under pressure, ranging from atmospheric pressure to very high pressures (up to 240 GPa). Orbital selective partial density of states of individual Fe-$ 3d $ and As-$ 4p $ orbitals near the Fermi level and their contributions in electronic band structures are affected differently due to applied pressure. At atmospheric pressure only Fe-$ 3d $ orbitals contribute significantly in the resulting electronic structure. But at higher pressures relatively larger presence of As-$ p $ orbital character bands near the FL suggest a larger overlap between Fe-$3d$ and As-$4p$ orbitals. Here we also recognize a metal-semiconductor transition in LiFeAs at 60 GPa pressure with band-gap around 0.16 eV, in accordance with the earlier studies \cite{111_hps_PCCP}. The inner electron FS are on the verge of Lifshitz transition in LiFeAs, so the nesting between hole and electron pockets are relatively weak. Occurrence of Lifshitz like topological transition for NaFeAs at 240 GPa pressure is visible. Most importantly, we found that at atmospheric pressure $ d_{xz} $ and $ d_{yz} $-orbitals are degenerate in crystal field splitting, which is lifted by the application of hydrostatic pressure. Increasing hybridization between the Fe-$ 3d $ orbitals with the surrounding As ligand's $ 4p $ orbitals at higher pressures may be responsible for this degeneracy lifting phenomena. We establish the above through estimation of $ d-p $ hybridization contribution in the full crystal field splitting energy, using maximally localized Wannier functions based formalism. The purely electrostatic energy contribution into full crystal field splitting also decreases with the increasing pressure, \textit{i.e.,} favoring orbital overlap. For LiFeAs intra-orbital and inter-orbital nearest neighbour hopping amplitudes are maximum at the semiconducting state and for metallic state they increase with increasing pressure. In case of NaFeAs, the intra-orbital nearest neighbour hopping amplitudes are maximum at relatively lower pressure (upto 40 GPa), while the inter-orbital hopping amplitudes becomes predominant at very high pressure. We also contemplate that the nearest-neighbour hopping amplitudes are more than one order of magnitude larger as compared to next-nearest-neighbour hopping amplitudes. Furthermore, the so-obtained tight binding parameters can be used to build simplified model Hamiltonians. These in turn allow access to the temperature dependent properties that are not accessible through density functional theory formalism. Finally, we believe our study will stimulate more theoretical and experimental activities in this domain.

\section*{CRediT authorship contribution statement}
The problem is formulated and conceptualized by H. Ghosh and S. Ghosh. All the theoretical calculations in the paper are carried out by S. Ghosh. The first draft of the manuscript was written by S. Ghosh and reformulated by H. Ghosh. Formal and critical analysis of the results were performed by both the authors.

\section*{Acknowledgements}
SG thanks A. Ghosh and A. Pokhriyal for fruitful discussions. Authors acknowledge Computer Division, RRCAT for providing scientific computing facilities. Financial support for this work to SG is provided by the HBNI-RRCAT. 

\section*{Author Declaration}
The authors declare no conflict of interest.

\section*{Data Availability Statement}
The data that support the findings of this study are available from the corresponding author upon reasonable request.

\bibliographystyle{apsrev4-1}   
\bibliography{MS4ref}
\end{document}